\documentclass[12pt]{article}
\usepackage[english]{babel}
\RequirePackage[colorlinks]{hyperref}
\usepackage{setspace}
\usepackage{booktabs}
\usepackage{graphicx}
\usepackage{rotating}
\usepackage{hyperref}
\usepackage{pdfsync}
\usepackage[margin=2cm,a4paper]{geometry}
\usepackage{amsmath,amsfonts,bm,amssymb} 
\usepackage{amsthm}

\newcommand{\ud}{\mathrm{d}}

\newcommand{\pval}{p_\mathrm{value}}

\begin{document}

\title{Modelling systemic price cojumps with Hawkes factor models}

\author{Giacomo Bormetti$^{\textrm{a,b,}}$\footnote{These authors contributed equally to this work.}~~\footnote{Corresponding author: giacomo.bormetti@sns.it, phone +39 050509159}~, Lucio Maria Calcagnile$^{\textrm{a,b,}*}$,\\ 
Michele Treccani$^{\textrm{c,b,}*}$, Fulvio Corsi$^\textrm{a}$, Stefano Marmi$^\textrm{a,d,b}$, Fabrizio Lillo$^\textrm{a,e,f,b}$}

\date{\today}
\maketitle
\small
\begin{center}
  $^\textrm{a}$~\emph{Scuola Normale Superiore, Piazza dei Cavalieri 7, Pisa, 56126, Italy}\\
  $^\textrm{b}$~\emph{QUANTLab\hspace{2pt}\footnote{www.quantlab.it}, via Pietrasantina 123, Pisa, 56122, Italy}\\
  $^\textrm{c}$~\emph{LIST S.p.A., via Pietrasantina 123, Pisa, 56122, Italy}\\
  $^\textrm{d}$~\emph{CNRS UMI 3483 - Laboratorio Fibonacci, Piazza dei Cavalieri 7, Pisa, 56126, Italy}\\
  $^\textrm{e}$~\emph{Dipartimento di Fisica e Chimica, Universit\`a degli Studi di Palermo, Viale delle Science Ed. 18, Palermo, 90128, Italy}\\
  $^\textrm{f}$~\emph{Santa Fe Institute, 1399 Hyde Park Road, Santa Fe, NM 87501, USA}\\
\end{center}
\normalsize

\vspace{0.3cm}

\smallskip

\begin{abstract}
  Instabilities in the price dynamics of a large number of financial assets are a clear sign of systemic events. By investigating a set of 20 high cap stocks traded at the Italian Stock Exchange, we find that there is a large number of high frequency cojumps. We show that the dynamics of these jumps is described neither by a multivariate Poisson nor by a multivariate Hawkes model. We introduce a Hawkes one factor model which is able to capture simultaneously the time clustering of jumps and the high synchronization of jumps across assets.  
\end{abstract}

\noindent\textbf{JEL:} C32, C5, C51, G01, G10\\
\textbf{Keywords:} cojumps, Hawkes processes, systemic shocks, high frequency data

\newpage

\section{Introduction}

Modelling the dynamics of security prices is of paramount importance for risk control, derivative pricing, trading, and to understand the behaviour of markets. The most popular class of stochastic models for security prices is certainly the class of jump diffusion models, where the price is described by a continuous diffusive part plus a jump component, modelling discontinuities in asset prices. In the majority of models the jump component is described by a compound Poisson process. In a multi-asset setting the classical jump diffusion models assume that the jump component of different assets are independent Poisson processes and the only dependence is possibly in the size of the jumps. An incomplete list of recent studies on jumps following this approach includes theoretical work on nonparametric jump identification \cite{Ait04,BarShe06,JiaOom08,LeeMyk08,AitJac09,LeeHan10,Corsi_etal:2010}, as well as empirical
analysis \cite{BolLawTau08,MahMcc04,Boletal09,AndBolDob07,DasUpp04,WriZho07}, and applications to asset pricing \cite{DufPanSin00,EraJohPol03,Era04,BolTodLi12,BolTod11}. General reviews on jumps are  \cite{ConTan04} and \cite{BarShe07}. However, there has been less research on cojumps (i.e. simultaneous jumps in two or more stock prices) 
and most of them are of empirical nature such as \cite{DunMcKSmit09, LahLauNee11, DunHvo12, GilShaTay12}; exceptions are the papers of \cite{GobMan07,JacTod09} and \cite{AitCacLae12} on cojump estimation and modelling.

Poisson jump processes have the great advantage of being analytically tractable. However one might argue that the jump component of the price process could be described instead by a process where jumps are clustered in time. The same objection emerges when one considers many assets.  Markets are nowadays more and more interconnected and it is a priori reasonable to expect that some sort of synchronization between the jumping times of different asset is present.

This synchronization effect had its most spectacular appearance during the May 6, 2010 Flash Crash. According to the SEC-CFTC, the crash started from a rapid price decline 
in the E-Mini S\&P 500 market~\cite{sec,Kyle}. However in a very short time the price drop propagated towards ETFs, stock indices and their components, and derivatives. 
For example, the Dow Jones Industrial Average plunged about nine percent, only to recover those losses within minutes. The contagion effect can be extremely rapid in liquid 
markets~\cite{Nanex,Johnson} and leads to a strongly synchronized discontinuous movement of the price of many assets. This type of systemic events can not be described by a model where price jumps follow independent processes. 

In this paper we show that indeed the dynamics of jumps of a portfolio of stocks deviates significantly from a collection of independent Poisson processes. 
The deviation that we observe is twofold. On one side, by considering individual assets, we find evidence of time clustering of jumps, clearly inconsistent with a Poisson process. 
This means that the intensity of the point process describing jumps depends on the past history of jumps, and a recent jump increases the probability that another jump occurs. 

The second deviation from the Poisson model is probably more important, especially in a systemic context. We find a strong evidence of a high level of synchronization between the jumping times of a portfolio of stocks. In other words, we find a large number of instances where several stocks (up to 20, the size of our set) jump at the same time (by using a resolution of a minute). This evidence is absolutely incompatible with the hypothesis of independence of the jump processes across assets.

Taking together these empirical deviations from the independent Poisson model, there is a need of a suitable modelling of the multi-asset jump process and this is the main methodological contribution of this paper. In order to model the time clustering of jumps for individual assets we propose the use of a class of self-exciting point processes, termed Hawkes processes. These processes were introduced more than forty years ago~\cite{Hawkes:1971}, and have been widely employed to model earthquake data~\cite{Vere-Jones:1970,Vere-Jones_Ozaki:1982,Ogata:1988}. 
In the last years, Hawkes processes have experienced an increasing popularity in mathematical finance and econometrics. One of the first applications to financial time series is due to~\cite{Bowsher:2007}, and a wide literature review in this context is  collected in~\cite{Bauwens_Hautsch:2009}. Among more recent developments not covered by the latter reference we mention~\cite{Muni-Toke:2011} where these processes are applied to the order flow in a continuous double auction market, \cite{Muni-Toke_Pomponio:2012} for the modelling of trades-through 
orders in a limit order book, \cite{Bacry_etal:2013} where Hawkes processes are used to introduce a new stochastic model for the variation of tick-by-tick asset price both in one and two dimension able to reproduce the strong microscopic mean reversion and the Epps effect, and \cite{Filimonov_Sornette:2012}, which introduces
a measure of the market activity providing a direct access to the level of its endogeneity and as a potential predictor of market micro instabilities (a critical review of this paper is given in~\cite{Hardiman_etal:2013}).
  
In this paper we use Hawkes processes for modelling the dynamics of jumps of individual assets and we show that they describe well the time clustering of jumps. However the direct extension of the application of Hawkes processes to describe the dynamics of jumps in a multi-asset framework is highly problematic and inconsistent with data. In fact, from a methodological point of view, even by using a simple two parameter kernel (for example, exponential) of the process, the number of parameters to estimate a Hawkes process on $N$ assets is $N(2N+1)\sim O(N^2)$, which is clearly too high. Moreover, even when we consider $N=2$ stocks, we empirically find that a bivariate Hawkes model is unable to describe the empirical data, especially to replicate the high number of {\it synchronous} jumps that we observe. This is due to the fact that the kernel structure of Hawkes is more suited to model lagged jumps rather than synchronous jumps. 

For this reason, the main methodological contribution of this paper is the introduction of Hawkes factor models to describe systemic cojumps. We postulate the presence of an unobservable point process describing a market factor. When this factor jumps, each asset jumps with a given probability, which is different for each stock. In general, an asset can jump also by following an idiosyncratic point process. In order to capture also the time clustering of jumps, we model the point processes as Hawkes processes. We show how to estimate this model and discriminate between systemic and idiosyncratic jumps. We show that the model is able to reproduce both the longitudinal and the cross sectional properties of the multi-asset jump process. 

Recent approaches sharing some aspects with the current paper are discussed in~\cite{AitCacLae12,Ait-Sahalia_Hurd:2012} and~\cite{GilShaTay12}. 
The former design a model of asset returns able to capture periods of crisis characterized by contagion and consider it to solve the 
problem of optimal investment-consumption for a log-utility investor. The jump diffusion component of the dynamics is described in terms of a class of multi-dimensional 
Hawkes models, and the authors discuss an estimation methodology based  on the Generalized Method of Moments. Not surprisingly, when they estimate the model on real data,
they face the problem of the curse of dimensionality and limit most of the study to the two asset case. Our one factor approach solves the
calibration issue in a natural way and therefore represents a viable alternative to their model.
By using data sampled at a frequency of eleven minutes the authors of~\cite{GilShaTay12} find empirical support to the hypothesis that stocks tend to be involved in 
systematic cojumps\hspace{2pt}\footnote{From a terminological viewpoint, authors of~\cite{GilShaTay12} define a systematic cojump as an instance when one stock and a market index jump 
at the same time. In this paper we use the term systemic cojump to indicate a sizeable number of stocks jumping simultaneously. In Section~\ref{sec:factor-model} we show how to identify systemic cojumps in a self consistent way.}, rejecting an assumption of independence in jump arrival times among stocks. We find strong evidence of this result also at the frequency of one minute and furthermore we extensively investigate the properties of self and mutual excitation possessed by the jumps series of different stocks.

The paper is organized as follows. In Section 2 we present our dataset and in Section 3 we summarize our jump detection method (detailed in Appendices). The identification of jumps is a delicate topic and therefore we are careful in using a robust identification method in order to minimize the number of false positives. In Section 4  we provide empirical evidence of the large number of systemic cojumps and we describe some simple statistical properties. In Section 5 we present the statistics we use to test our models in their ability of reproducing the multiple jumps of a single stock in a given time window and the cross jumps, i.e.~the occurrences of jumps in two different stocks in a given time window. Section 6 discusses how Hawkes processes fit the jump process in a univariate and in a multivariate setting. In Section 7 we present the Hawkes factor model approach, showing how to estimate it and the results obtained in the investigated dataset. Finally, in Section 8 we draw some conclusions.

\section{Data description and data handling}\label{sec:data_handling}

The analyses reported in this work are performed on tick-by-tick transaction data for the period from 5th March 2012 to 9th July 2012 
that have been made available to us by LIST S.p.A.\hspace{2pt}\footnote{www.list-group.com}. 
In this work we investigate 20 among the most liquid stocks of the FTSE MIB index of the Italian stock exchange Borsa Italiana. We sample prices at a frequency of one minute, taking the last executed price, to obtain 1-minute logarithmic returns. The total number of days in the period is 88, with 505 intraday returns each day.

In Table~\ref{table:stocks} we list the twenty stocks, along with the ISIN code and the average number of trades per day. From the last column we see that the assets are characterized by a certain level of heterogeneity, 
since the trade activity varies between 2.8 and 27.4 thousand transactions per day. A low level of activity implies in general a higher probability for the absence of transactions
inside a given sampling interval, and this effect has important implications for the methodology that we use for the detection of jumps.
\begin{table}
  \centering
  \begin{tabular}{|l|l|r|}
    \hline
    Company & ISIN code & trades per day ($\times 10^{3}$)\\
    \hline \hline
    Assicurazioni Generali & IT0000062072 & 7.5\\
    Mediobanca & IT0000062957 & 3.6\\
    Banca Popolare di Milano & IT0000064482 & 4.9\\
    Saipem & IT0000068525 & 5.4\\
    Intesa Sanpaolo & IT0000072618 & 18.1\\
    Mediaset & IT0001063210 & 3.1\\
    Banca Monte dei Paschi di Siena & IT0001334587 & 9.2\\
    Fiat & IT0001976403 & 9.6\\
    Enel & IT0003128367 & 10.7\\
    Eni & IT0003132476 & 10.4\\
    UBI Banca & IT0003487029 & 3.0\\
    Telecom Italia & IT0003497168 & 6.3\\
    Finmeccanica & IT0003856405 & 4.7\\
    Prysmian & IT0004176001 & 2.8\\
    Banco Popolare & IT0004231566 & 4.2\\
    Pirelli \& C. & IT0004623051 & 4.6\\
    Fiat Industrial & IT0004644743 & 5.1\\
    UniCredit & IT0004781412 & 27.4\\
    Tenaris S.A. & LU0156801721 & 3.3\\
    STMicroelectronics N.V. & NL0000226223 & 4.3\\
    \hline
  \end{tabular}
  \caption{List of the twenty investigated stocks, with the ISIN code and the average number of transactions per day.}
  \label{table:stocks}
\end{table}

The process from tick-by-tick data to jump identification can be summarized as follows. 
First, anomalous values in tick-by-tick price data are detected and removed. The algorithm for the outliers detection that we use is due to 
Brownlees and Gallo~\cite{Brownlees_Gallo:2006} and it is explained in Section~\ref{sub:brownlees_gallo} of the Appendix.
We find no outliers at all for 13 stocks and few units for the other 7, for the vast majority concentrated on the very first minute of the day. 
Removing outliers can in principle introduce a distortion in the analysis. However, the number of anomalous prices that we identify 
is extremely low, and the probability of an actual distortion of returns, computed at the minute level, is negligible.
Cleaned prices are then sampled at a frequency of one minute and logarithmic returns are calculated carrying forward to the next sampling instant 
the last price observed within the sampling window.

In the identification of jumps, an important care should be devoted to the way in which intervals without trades are treated. In fact, jump detection methods typically compare returns with local volatility. In periods of low liquidity (or missing observations) one risks to underestimate volatility and to identify relatively small price returns as jumps. There are two different situations which lead to having no price observations in a given time interval. 

The first one occurs in correspondence of volatility auction phases.
According to the rules of the market~\cite{Rules:2012,Instructions:2012}, whenever the price exits a reference interval, reaching too high or too low levels, the continuous double auction phase of the exchange is suspended and a volatility auction starts. In such a phase we consider the related returns as \textit{not available}. We do not have direct information on volatility auctions, but we infer their presence from market data. Since the investigated stocks are among the most liquid ones, we are reasonably sure to be able to distinguish volatility auction phases from continuous double auction phases of low liquidity. For more details, see the discussion in Section~\ref{sec:vola_auction} of the Appendix.

The second mechanism for missing price observations is when the stock is available for trades, but still there are no transactions in that particular minute. 
There are several ways to treat these cases. To fix the notation, let $i$ and $i+1$ label two consecutive sampling instants between which no trade has been made.
The most common way to deal with the missing observation of the price between times $i$ and $i+1$ is to bring forward the last recorded price. 
This means setting the price $p_{i+1} = p_i$ and the log-return $r_{i+1} = 0$. Sampling intervals with no transactions in them are therefore given a zero return. This is our first method of dealing with Missing Observations, which we call MO1. 
 An alternative method of treating minutes with no transactions is simply to consider the corresponding returns as not available data 
($r_{i+1} = \texttt{NA}$), thus avoiding to give them a numerical value. The next available return will be at the first sampling time after the next available observation. 
It can be defined as the usual difference of the sampled logarithmic prices after and before the missing observations (method MO2) or as this quantity rescaled by the square root of the time between the two sampling times (MO3). For instance, for the price series $p_0,p_1,-,-,p_4,p_5$ we construct the following three returns series:
\begin{itemize}
  \centering
  \item[\textbf{MO1}] \qquad $\log \frac{p_1}{p_0}, \phantom{1}0, \phantom{1}0, \phantom{\frac{1}{\sqrt{3}}} \log \frac{p_4}{p_1}, \log \frac{p_5}{p_4}$;
  \item[\textbf{MO2}] \qquad $\log \frac{p_1}{p_0}, \texttt{NA}, \texttt{NA}, \phantom{\frac{1}{\sqrt{3}}} \log \frac{p_4}{p_1}, \log \frac{p_5}{p_4}$;
  \item[\textbf{MO3}] \qquad $\log \frac{p_1}{p_0}, \texttt{NA}, \texttt{NA}, \frac{1}{\sqrt{3}} \log \frac{p_4}{p_1}, \log \frac{p_5}{p_4}$.
\end{itemize}
The reason for the rescaling of the return in method MO3 is that we do not want to identify as a jump a price change that is compatible with a diffusive 
random movement of the efficient price.

Once the returns are defined according to one of the three possible methodologies discussed above, they are checked in order 
to eliminate possible large values due to stock splits or merges. We check for the presence of stock splits in our data through the detection of returns greater 
than 0.2 in absolute value. This would detect, for example, a 3-for-2 split or a 4-for-5 merge. In our data we do not find any such return.

The last step to perform is the removal of daily seasonalities, by filtering out the intraday pattern from raw returns. As it is well known, intraday returns show significant seasonal behaviour, as the dynamics of markets is greatly variable during the day. Opening and closing periods generally show a higher volatility than the rest of the day, since traders are more active during these phases. If these daily seasonalities are not properly filtered out, spurious jump detection may happen in correspondence of these periods. We describe the details of how we deal with this issue in Section~\ref{sec:intraday_pattern}.

\section{Jump identification}\label{sec:jump_identification}

After performing the data cleaning procedures explained in Section~\ref{sec:data_handling} and detailed in Appendix~\ref{app:data_cleaning}, we proceed with the detection of jumps. 
Following a standard approach, we estimate the local volatility $\sigma$ and then test whether the ratio between absolute returns and local volatility is above a given threshold, that is, if
\[ \frac{|r|}{\sigma} > \theta . \]
In this work we take $\theta = 4$ as in~\cite{Joulin_etal:2008} and return series are obtained from historical prices as reported in Section~\ref{sec:data_handling}. In order to estimate the local volatility, we can follow several strategies available in literature. We base our approach on the realized absolute variation and the realized bipower variation, whose asymptotic theory is treated for example in~\cite{Barndorff-Nielsen_Shephard:2004}. Our estimators of local volatility based on these quantities are defined by the exponentially weighted moving averages
\begin{equation} \label{eq:sigma_abs}
	\hat{\sigma}_{\text{abs},t} = \mu_1^{-1} \alpha \sum_{i > 0} (1-\alpha)^{i-1} |r_{t-i}|
\end{equation}
and
\begin{equation} \label{eq:sigma_bv}
	\hat{\sigma}_{\text{bv},t}^2 = \mu_1^{-2} \alpha \sum_{i > 0} (1-\alpha)^{i-1} |r_{t-i}| |r_{t-i-1}| ,
\end{equation}
where $\mu_1 = \sqrt{\frac{2}{\pi}} \simeq 0.797885$ and the parameter of the exponential averages is $\alpha = \frac{2}{M+1}$, with $M = 60$, which corresponds to a half-life time of nearly 21 minutes. In order to avoid biased estimates due to the presence of jumps in past returns, we actually use modified versions of the estimators~(\ref{eq:sigma_abs}) and~(\ref{eq:sigma_bv}) that use only returns where no jump is detected. 
In fact the presence of a jump in the estimation window leads to an overestimation of the volatility.
We refer the reader to the Appendix~\ref{sec:vol_proxy} for the details on the implementation of our estimators and the asymptotic properties of the realized absolute variation and the realized bipower variation.

Since we construct three series of returns (which differ only in correspondence and proximity of missing observations), for each one we obtain 
the volatility estimates $\hat{\sigma}_\text{abs}$ and $\hat{\sigma}_\text{bv}$, we have six volatility-normalised series $\big \{ \frac{r_t}{\sigma_t} \big \}$, each of which with its own set of detected jumps. In order not to be sensitive to the choice of how to treat missing observations nor to the choice of volatility estimation, we take as final series of jumps the intersection of the six sets of jumps. This should also minimize spurious jump detection, as suggested in~\cite{Dumitru_Urga:2011}.

\subsection{Performance of the jump detection methods on simulated data}

In the next section we discuss the results that we have found about identified jumps. 
Preliminarily, we conduct a simulation study to assess the effectiveness of our entire procedure from data cleaning to jump detection. 
The usual approach is to assume the observed log-price $p(t)$\footnote{In this paper we express the time dependence among parenthesis when we are dealing with continuous time variables, whilst we use indices for discrete time models and sampled data.} as coming from the sum
\begin{equation} \label{eq:observed_price}
  p(t) = X(t) + \epsilon(t)
\end{equation}
of an efficient log-price $X(t)$, following a simple jump diffusion model
\begin{equation} \label{eq:jump-diffusion_model}
  \ud X(t) = \mu(t) \, \ud t + \sigma(t) \, \ud W(t) + \kappa(t) \, \ud J(t) ,
\end{equation}
where the three terms on the right hand side are the drift, diffusion and jump components, and a microstructure noise $\epsilon_t \sim \text{i.i.d. } \mathcal{N} (0,\eta^2)$. 
Following Fan and Wang~\cite{Fan_Wang:2007}, for the dynamics of the volatility $\sigma(t)$ we take the Geometric Ornstein-Uhlenbeck model
\begin{equation}
  \ud \log \sigma^2(t) = - (0.6802 + 0.1 \log \sigma^2(t)) \, \ud t + 0.25 \, \ud W^\prime(t)\,,
  \label{eq:GOUvola}
\end{equation}
with correlation between the Wiener processes $W(t)$ and $W^\prime(t)$ equal to -0.62.
Simulating model of Equations~(\ref{eq:observed_price}) and~(\ref{eq:jump-diffusion_model}) we fix the drift term $\mu$ equal to 0, and the microstructure noise standard deviation 
$\eta$ to $10^{-5}\ud t$.
As far as the process $J(t)$ is concerned, we set its rate equal to three jumps per day, and the size $\kappa$ as a multiple of the spot volatility bootstrapped  
from the sample of the empirical data. Using the Euler discretization scheme we obtain simulated high-frequency log-prices from the model at a frequency 
of one minute, which corresponds to the sampling frequency of our data. The time convention that we adopt fixes the one day horizon equal to one, so we have $\ud t = 1 / 1440$. 
We simulate data for a period 50 times longer than that of the sample data, that is, 88 $\times$ 50 = 4400 days. We preliminarily draw a Monte Carlo simulation of the 
process~(\ref{eq:GOUvola}) and then we rescale it by adding the intraday volatility pattern. The seasonal volatility is used in~(\ref{eq:jump-diffusion_model}) 
to generate one realization of the process $X(t)$.
Finally, for each day we discard some observations in order to reproduce the observed intertrade times, thus introducing artificial missing observations to investigate the role of 
methods MO1, MO2, MO3 of Section~\ref{sec:data_handling} in the jump detection. This operation is performed through a random sampling 
from the empirical distribution of the intertrade times.

Table~\ref{tab:size_power_tests} shows the performance of our jump detection procedure for simulations calibrated on two stocks: Intesa Sanpaolo, the second most liquid stock among our data, and Banco Popolare, which has a much lower liquidity. The performance is presented in terms of size and power of the test, 
for the jump series coming from methods MO1, MO2, MO3, with the two different estimates of volatility $\hat{\sigma}_{\text{abs}}$ and $\hat{\sigma}_{\text{bv}}$. 
In the last column and row we report  the intersections between the different methods.
Simulations calibrated on Intesa Sanpaolo show small differences among the three methods of dealing with the missing observations in sampled data. This is expected, since there are relatively few points in which the corresponding return series differ. Also the intersection of jumps detected in methods MO1, MO2, MO3 has performances comparable to those of each of the three methods.

A benchmark simulation calibrated on the stock Banco Popolare shows instead a very different picture, namely it highlights the significance of the way of filling the missing observations for a stock of low liquidity. If such missing observations are filled with zeroes (method MO1), volatility estimates from past returns are lower than for methods MO2 and MO3. This translates into detecting many more jumps, both true and spurious. In absolute terms, average counts of right and false positives are respectively 189 and 143 for method MO1, 146 and 45 for method MO2, 147 and 10 for method MO3, 134 and 7 for the intersection of all methods. Thus, results of this simulation clearly show that taking the intersection of jumps detected in methods MO1, MO2, MO3 drastically reduces the number of false positives, although this is accompanied also with a significant reduction of right positives. Taking a conservative point of view, we are ready to miss the detection of some true jumps as long as this procedure allows minimizing false positive hits. This simulation also demonstrates in a very clear way how the method MO3 is effective in detecting much fewer false positives than method MO2, still detecting the same amount of true jumps.

\begin{table}
  \begin{center}
    \begin{tabular}{|c|c|c|c|c|c|c|c|c|}
      \hline
      \rule[-0pt]{0pt}{13pt}      & \multicolumn{2}{c|}{MO1} & \multicolumn{2}{c|}{MO2} & \multicolumn{2}{c}{MO3} & \multicolumn{2}{|c|}{$\bigcap_{\text{MO1},\text{MO2},\text{MO3}}$}\\
      \rule[-5pt]{0pt}{5pt}      & Size & Power & Size & Power & Size & Power & Size & Power \\
      \hline
      \rule[-5pt]{0pt}{19pt} $\sigma_\text{abs}$              & 0.036\% & 63.1\% & 0.033\% & 62.5\% & 0.029\% & 62.7\% & 0.028\% & 62.1\%\\
      \hline
      \rule[-5pt]{0pt}{19pt} $\sigma_\text{bv}$               & 0.039\% & 63.4\% & 0.037\% & 62.7\% & 0.033\% & 62.7\% & 0.032\% & 62.3\%\\
      \hline
      \rule[-5pt]{0pt}{19pt} $\bigcap_{\text{abs},\text{bv}}$ & 0.031\% & 61.6\% & 0.029\% & 61.0\% & 0.026\% & 61.0\% & 0.025\% & 60.6\%\\
      \hline
    \end{tabular}

    \begin{tabular}{|c|c|c|c|c|c|c|c|c|}
      \hline
      \rule[-0pt]{0pt}{13pt}      & \multicolumn{2}{c|}{MO1} & \multicolumn{2}{c|}{MO2} & \multicolumn{2}{c}{MO3} & \multicolumn{2}{|c|}{$\bigcap_{\text{MO1},\text{MO2},\text{MO3}}$}\\
      \rule[-5pt]{0pt}{5pt}      & Size & Power & Size & Power & Size & Power & Size & Power \\
      \hline
      \rule[-5pt]{0pt}{19pt} $\sigma_\text{abs}$              & 0.369\% & 72.9\% & 0.110\% & 56.4\% & 0.027\% & 58.0\% & 0.019\% & 52.3\%\\
      \hline
      \rule[-5pt]{0pt}{19pt} $\sigma_\text{bv}$               & 0.390\% & 74.7\% & 0.135\% & 59.7\% & 0.031\% & 58.3\% & 0.025\% & 55.1\%\\
      \hline
      \rule[-5pt]{0pt}{19pt} $\bigcap_{\text{abs},\text{bv}}$ & 0.324\% & 71.7\% & 0.102\% & 55.2\% & 0.023\% & 55.8\% & 0.016\% & 50.9\%\\
      \hline
    \end{tabular}

    \caption{Size and power of the jump tests, determined on a simulation calibrated on real data of Intesa Sanpaolo (top) and Banco Popolare (bottom).
    The size is the ratio between the number of false positives and the number of minutes when the jump is not present. The power is one minus the ratio between the number of 
    right positives and the number of real jumps.
  }
    \label{tab:size_power_tests}
  \end{center}
\end{table}


\section{Basic statistics of jumps of individual stocks and of systemic jumps}

In this section we present some basic statistical properties of detected jumps. In particular we will consider the restricted set of jumps identified simultaneously by all methods, in order to minimize the number of false positives. We then consider some simple statistical characterization of cojumps, i.e.~instances in which the prices of at least two stocks jump in the same minute.

\subsection{Jumps of individual stocks}

In the previous sections we have introduced several methods to identify jumps in price time series. By using Monte Carlo simulations of a realistic model of price dynamics, we have shown also that different methods display different ability to identify jumps correctly, both in terms of false positives and in terms of false negatives. The use of the intersection among the different methods improves significantly the identification procedure, at least on simulated data.

Table~\ref{tab:numberofjumps_allmethods} shows the number of jumps for the stock Monte dei Paschi di Siena, detected with the six different methods. For this specific stock, the number of jumps ranges from 200 to 281, showing a significant dependence on the used method. The table shows also the size of the intersection of the sets of detected jumps. When one considers jumps detected by all methods the number of jumps falls to 178. This is the restricted set of events that we will consider. We notice that, in this case, by estimating the volatility as $\hat{\sigma}_\text{abs}$ we find more jumps than when using $\hat{\sigma}_\text{bv}$, regardless of what the method of treating missing observation is. However this is not always the case. For the stock Intesa Sanpaolo it is the other way round: more jumps are detected when we estimate local volatility by $\hat{\sigma}_\text{bv}$ than when we use $\hat{\sigma}_\text{abs}$, in methods MO1, MO2 and MO3. It is not clear to us whether there are features of the return series (such as quantity and position of the missing observations) which are systematically responsible for the first or the second scenario to happen.

\begin{table}[h]
	\centering
	\begin{tabular}{|c|ccc|c|}
		\hline
		& MO1 & MO2 & MO3 & $\bigcap_{\text{MO1,MO2,MO3}}$\\
		\hline
		$\hat{\sigma}_\text{abs}$ & 281 & 228 & 217 & 205\\
		$\hat{\sigma}_\text{bv}$ & 260 & 208 & 200 & 190\\
		\hline
		$\bigcap_{\text{abs,bv}}$ & 239 & 196 & 186 & \textbf{178}\\
		\hline
	\end{tabular}
	\caption{Summary table of the number of detected jumps by using different methods and their intersection. 
	The investigated stock is Monte dei Paschi di Siena.}
	\label{tab:numberofjumps_allmethods}
\end{table}

We extend this analysis to the whole set of 20 stocks. In Table~\ref{tab:jumps_results} we report simple statistics about the jumps identified on the 20 stocks by using the intersection of the six methods. The number of jumps detected in the 88 days of our sample varies between 59 and 188 across the twenty stocks (that is, between 0.67 and 2.14 jumps per day per stock), with an overall average value of 108, corresponding to 1.23 jumps per day per stock\hspace{2pt}\footnote{It is worth noticing that in a recent study~\cite{Joulin_etal:2008} authors found more than seven jumps per stock per day when investigating a relatively large set of US stocks. Even if the threshold used is the same as ours, i.e.~$\theta=4$, they use only one method,
they do not detail the data cleaning procedure, and they investigate a much larger universe of stocks.
These might be reasons for the difference in the typical number of detected jumps.}. 

Five stocks show a statistically significant difference in the number of positive and negative jumps (at 10\% significance level), when one assumes a null model in which stocks have the same probability of jumping up and down. In such a model, the number of jumps in either direction has a binomial distribution. Interestingly, in all five cases the number of negative jumps is greater than that of positive jumps. Statistically significant asymmetry in the jumping direction is also found in the overall jump count (at 1\% significance level). The twenty stocks have also variable proportions of ``single jumps'', that is, jumps that do not occur simultaneously to jumps of other stocks, with two having even a greater number of cojumps than single jumps. Indeed, the complementary proportion of cojumps attains relatively high values, among 16\% and 57\%, suggesting that the cojumps play a relevant role in jumps' behaviour and motivating further analyses in this respect.

\begin{table}[ht]
  \centering
  \begin{tabular}{|l||r||r|r||r|r|}
    \hline
    ISIN & jumps & jumps up & jumps down & single jumps & cojumps\\
    \hline
    \hline
    IT0000062072 & 103 &           48 (47\%) &           55 (53\%) &  53 (51\%) &  50 (49\%)\\
    IT0000062957 &  63 &           29 (46\%) &           34 (54\%) &  38 (60\%) &  25 (40\%)\\
    IT0000064482 & 121 &           60 (50\%) &           61 (50\%) &  97 (80\%) &  24 (20\%)\\
    IT0000068525 &  93 &           46 (49\%) &           47 (51\%) &  56 (60\%) &  37 (40\%)\\
    IT0000072618 & 127 &           67 (53\%) &           60 (47\%) &  55 (43\%) &  72 (57\%)\\
    IT0001063210 &  59 &           28 (47\%) &           31 (53\%) &  44 (75\%) &  15 (25\%)\\
    IT0001334587 & 178 &  \textbf{73} (41\%) & \textbf{105} (59\%) & 150 (84\%) &  28 (16\%)\\
    IT0001976403 & 123 &           61 (50\%) &           62 (50\%) &  76 (62\%) &  47 (38\%)\\
    IT0003128367 & 188 &  \textbf{81} (43\%) & \textbf{107} (57\%) & 107 (57\%) &  81 (43\%)\\
    IT0003132476 & 155 &  \textbf{66} (43\%) &  \textbf{89} (57\%) &  95 (61\%) &  60 (39\%)\\
    IT0003487029 &  70 &           28 (40\%) &           42 (60\%) &  41 (59\%) &  29 (41\%)\\
    IT0003497168 & 129 &           74 (57\%) &           55 (43\%) &  79 (61\%) &  50 (39\%)\\
    IT0003856405 &  95 &           50 (53\%) &           45 (47\%) &  74 (78\%) &  21 (22\%)\\
    IT0004176001 &  74 &           41 (55\%) &           33 (45\%) &  46 (62\%) &  28 (38\%)\\
    IT0004231566 & 103 &           50 (49\%) &           53 (51\%) &  72 (70\%) &  31 (30\%)\\
    IT0004623051 & 115 &  \textbf{47} (41\%) &  \textbf{68} (59\%) &  85 (74\%) &  30 (26\%)\\
    IT0004644743 & 100 &           51 (51\%) &           49 (49\%) &  65 (65\%) &  35 (35\%)\\
    IT0004781412 & 118 &  \textbf{49} (42\%) &  \textbf{69} (58\%) &  57 (48\%) &  61 (52\%)\\
    LU0156801721 &  59 &           27 (46\%) &           32 (54\%) &  32 (54\%) &  27 (46\%)\\
    NL0000226223 &  86 &           39 (45\%) &           47 (55\%) &  51 (59\%) &  35 (41\%)\\
    \hline
    \hline
    total & 2159 & \textbf{1015} & \textbf{1144} & &\\
    \hline
    \hline
    average & 108.0 & 50.8 (47\%) & 57.2 (53\%) & &\\
    \hline
  \end{tabular}
  \caption{Number of detected jumps for the twenty stocks, with direction and cojumping information. Bold values are inconsistent (at 10\% significance level for single stocks, at 1\% significance level for the total counts) with a null assumption of equal probability of jumping up and jumping down. Single jumps occur at a time when no other stock jumps, while cojumps occur simultaneously with jumps in other stocks.}
  \label{tab:jumps_results}
\end{table}

We investigate the distribution of jumps during the trading day. The left panel of Figure~\ref{intraday} shows the histogram of the time of the day when a jump occurs. The figure is obtained by including all the jumps of the 20 stocks, but we count only once a minute where multiple stocks jump simultaneously. We observe that there is no clear periodicity in the number of jumps, indicating that our intraday pattern removal is quite effective. Moreover, apart a spike observed at the beginning of the day, there is no evidence of minutes of the day when it is more likely that one stock jumps.

\subsection{Systemic jumps}\label{sec:systemic_jumps}

\begin{figure}[ht]
  \begin{center}
    \includegraphics[scale=0.8]{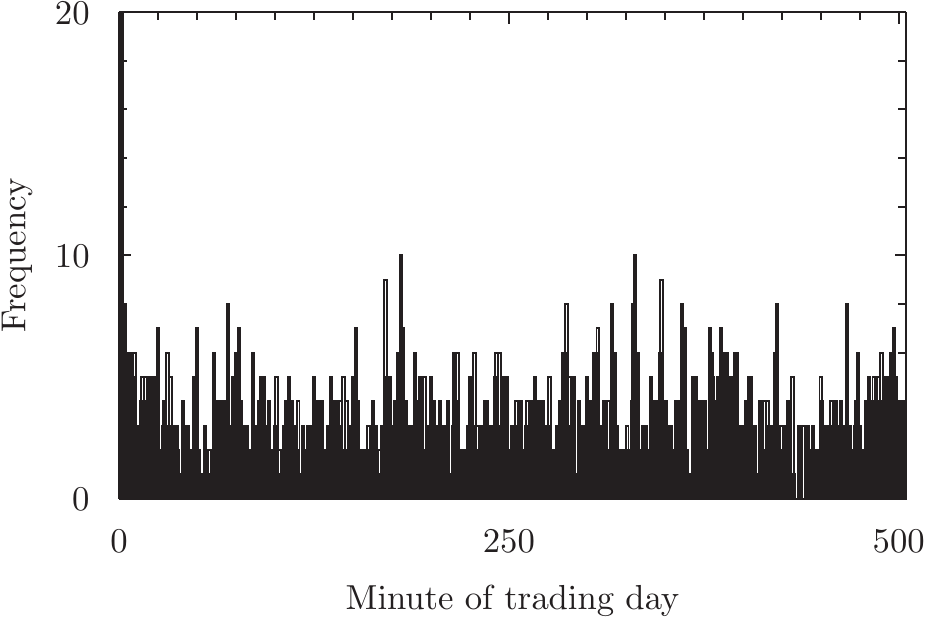}
    \includegraphics[scale=0.8]{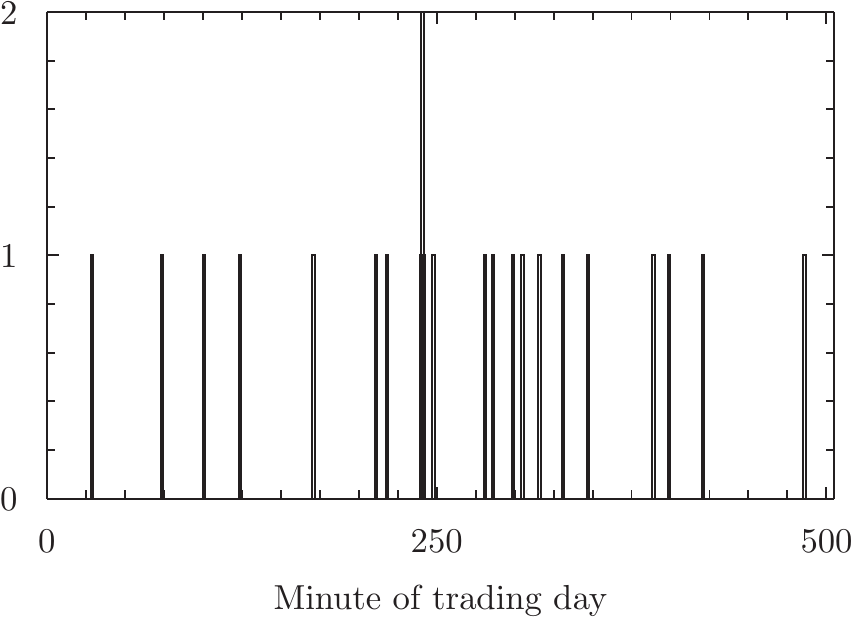}
    \caption{Left panel: intraday distribution of the time of the day when a jump occurs. The $x$ axis is the number of minutes from the beginning of the trading day. Notice that if more than one stock jump in a given minute, we count the minute only once in the histogram. This is done in order to avoid to count $N$ times an event when $N$ stocks jump simultaneously. Right panel: same figure as the left panel, but considering only cojumps of more than 5 stocks.}
    \label{intraday}
  \end{center}
\end{figure}

The main topic of this paper is systemic cojumps, i.e.~minutes when the price of a (possibly large) number of stocks displays a jump. For this reason it is important to investigate how frequently cojumps occur and to compare the observed statistics with those expected under some null hypotheses. 

\begin{figure}[ht]
  \begin{center}
    \includegraphics[scale=1.5]{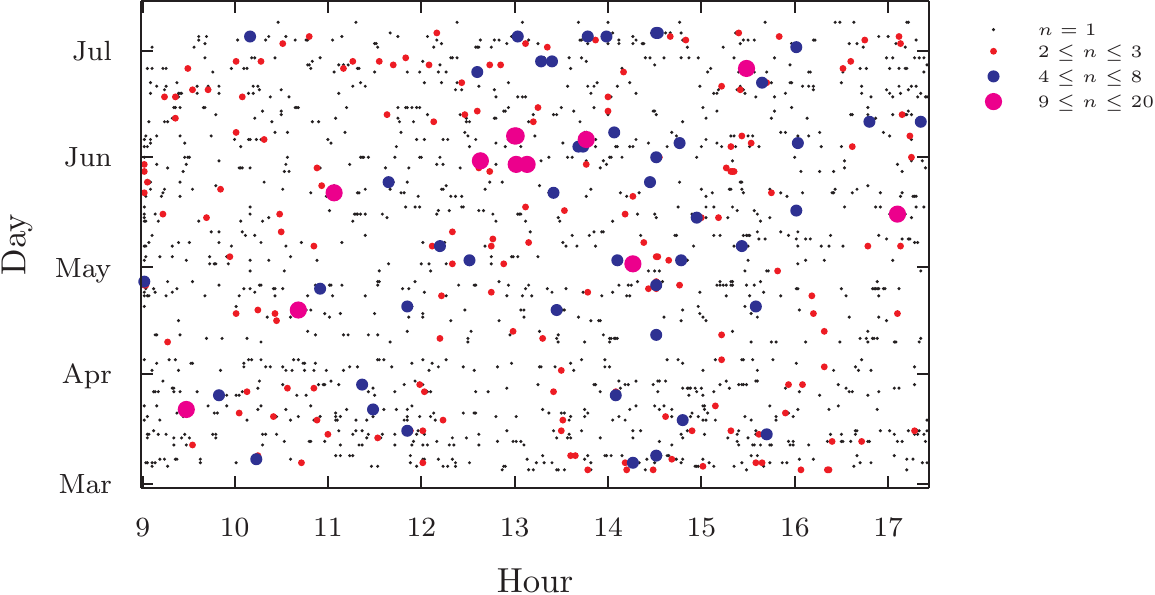}
    \caption{Time series of the cojumps observed in the set of 20 investigated stocks. The horizontal axis is the time of the day and the vertical axis is the day. 
    The size of the circle codes the number of stocks simultaneously jumping in a given minute.}
    \label{cojumps}
  \end{center}
\end{figure}

We start with some simple visualization of the occurrences of multiple cojumps. Figure~\ref{cojumps} shows the dynamics of the number of cojumps, indicating also the number of stocks that display a jump in a given minute. We notice that there are several occurrences where more than 8 stocks jump simultaneously (big circles). For example, we observe one case each when 10, 12, 13, 15, 16, 17, and 20 stocks jump simultaneously. Also the number of cases where more than 3 stocks jump (medium-sized circles) is quite high. For example, we observe 22 cases with 4 stocks, 15 cases with 5, 2 cases with 6, 4 cases with 7, 5 cases with 8, 3 with 9, and 2 with 11. Finally, there is a significant background of cases where two (136) or three (44) stocks jump simultaneously. Thus multiple stock cojumps are relatively frequent. 

By considering all the 240 events in which multiple stocks jump, we find only 7 cases in which not all the jumping stocks follow the same direction. There are 6 cases in which 2 stocks jump in opposite direction and 1 case in which two stocks jump down and one stock jumps up. The fact that when several stocks simultaneously jump they all move in the same direction suggests that a single common factor explains the jumping probability. In Section~\ref{sec:factor-model} we will develop this idea more formally by introducing models that capture this important feature of real data.

It is worth noticing that the Figure~\ref{cojumps} also indicates that there are no specific times of the day (for example corresponding to pre-announced news or opening of other markets) where the systemic cojumps are more frequent. This qualitative statement can be made more precise by drawing the histogram of the time of the day when a systemic cojump with more than 5 stocks jumping simultaneously (see right panel of Figure~\ref{intraday}). Therefore systemic jumps do not occur at preferential moments of the trading day.

\begin{figure}[ht]
  \begin{center}
    \includegraphics[scale=0.8]{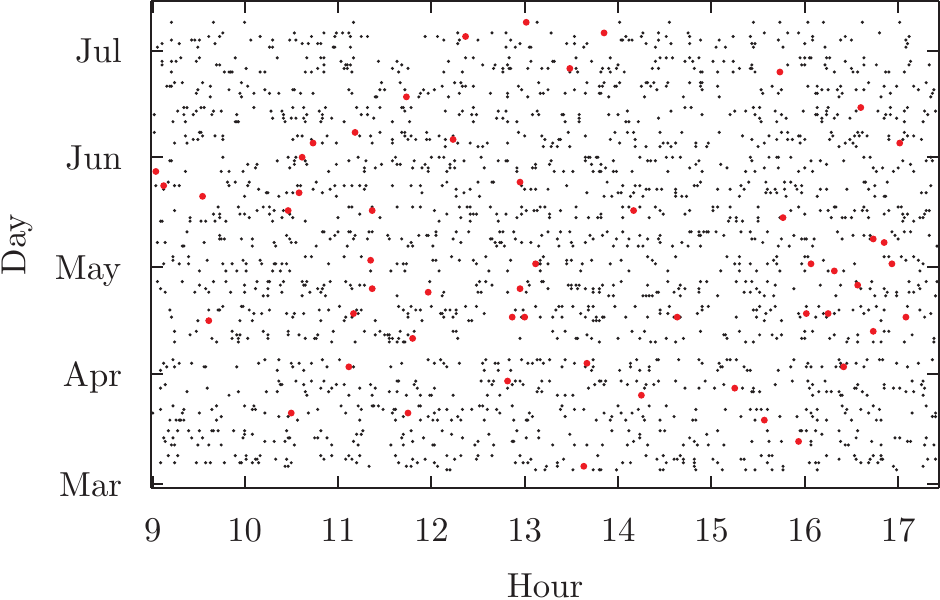}
    \includegraphics[scale=0.8]{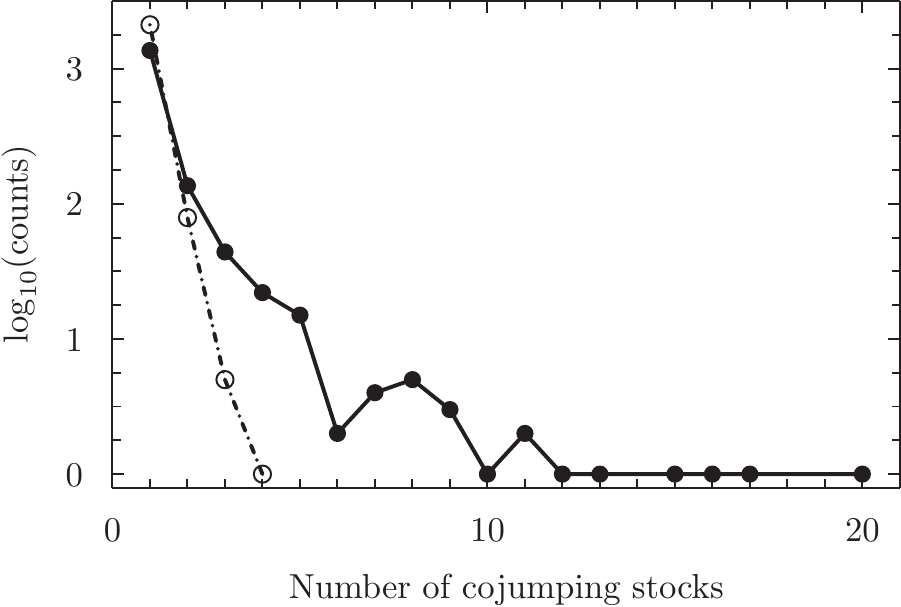}
    \caption{Left panel. Time series of the cojumps observed in a bootstrap replica of the real data preserving the jump intensity of each stock. 
      The horizontal axis is the time of the day and the vertical axis is the day. The size of the circle codes the number of stocks simultaneously jumping in a 
      given minute under the same convention of Figure~\ref{cojumps}. Right panel. Histogram of the number of stocks simultaneously cojumping in a minute (solid line). For comparison the dashed line shows the $0.01\%$ 
      confidence interval for the counts under the null hypothesis of independent but not identical Poisson processes.}
    \label{cojumpsNull}
  \end{center}
\end{figure}

In order to compare the pattern observed in Figure~\ref{cojumps} for the systemic cojumps with a null hypothesis, we perform a bootstrap analysis. Specifically, we construct a bootstrap replica independently for each stock. Therefore our replicas are consistent with a model of 20 independent but not identical Poisson processes for the jumps. In the next section we will use this model as the simplest benchmark model. In the left panel of Figure~\ref{cojumpsNull} we show the analogous of Figure~\ref{cojumps} for one of the bootstrap replicas of the real data. As it can be seen, in this replica there are no cases where more than three stocks jump simultaneously, a result that is clearly inconsistent with the real data. To be more quantitative, in the right panel of Figure~\ref{cojumpsNull} we show the histogram of the number of stocks jumping in a minute (in which at least one stock jumps) for real data (solid line). We compare it with the curve obtained by taking the $0.01\%$ confidence interval from bootstrap analysis. It can be seen that already the observed number of cojumps of two stocks is incompatible with the one observed in bootstrap test at the $0.01\%$ confidence. Moreover in $10,000$ bootstrap replicas we never observe a cojump with more than 4 stocks (observed only once) while in real data we have several cases with the price of many stocks simultaneously jumping. 

In conclusion, our descriptive analysis shows that (i) jumps are relatively frequent, even when one considers relatively strict detection criteria, (ii) there is a large number of cojumps, i.e.~minutes when a sizable number of stocks (up to 20!)~simultaneously jump, (iii) these cojumps show no clear timing inside the day, and, more important,  (iv) they are absolutely not compatible with a null model of independent but not identical Poisson processes. In Section~\ref{sec:factor-model} we will introduce models able to describe the jump dynamics of a set of stocks.

\section{A multi-scale statistical test based on multiple jumps and cross jumps detection}

The empirical observation of a large number of cojumps requires a rigorous statistical test to compare the observed behaviour of jumps with the prediction of  multivariate point processes. The main problem in working with jumps is that their number is relatively small, and therefore the statistics must be suited to work on small samples. Moreover, we want to use a multi-scale statistical test, i.e.~a test that is able to identify deviations from models at different time scales.
If the observation time is much larger than the correlation time of the process and we only focus on the counting of events disregarding the interarrival times,
then a correlated point process is undistinguishable from a Poisson process. Therefore we use a test that considers simultaneously different time scales. Finally, even if the main topic of this paper is cojumps of different stocks, we are also interested in investigating deviation from a Poisson behaviour at the individual stock level. Therefore we shall introduce two related statistics, one for consecutive jumps of the same asset and one for jumps of different stocks occurring at close times. 

Since the distributional and time correlation analysis is unfeasible with small samples of jumps we will consider the following statistics that measures the frequency of multiple jumps occurring in the same time windows.

Specifically, when considering individual stocks, we define a {\it multiple jump} (MJ) as an event which occurs when 
\textit{at least two jumps of the same stock price} are observed inside a time window of a fixed length $w$. Let us call $s_i$ the number of jumps inside the $i$-th window. An estimator of the MJ probability in a window of length $w$ over a sampling period of length $N$ is given by
\begin{equation}
  \hat{p}_w^\text{MJ} = \frac{\sum_{i = 1}^{\left \lfloor \frac{N}{w} \right \rfloor} 
  \mathbf{1}_{s_{i} \geq 2 }}{\left \lfloor \frac{N}{w} \right \rfloor}\,,
  \label{eq:psw}
\end{equation}
where $\mathbf{1}_A$ is the indicator function of the event $A$, and the symbol $\lfloor N/w\rfloor$ corresponds to the integer part of the ratio $N/w$.

The same idea can be extended to capture also the cross sectional clustering of jumps, in particular the evidence of a large number of simultaneous jumps of different stocks. The second notion which we will work with is therefore that of {\it cross jumps} (CJ) between two stocks defined when \textit{both stocks jump at least once}  inside a given time window of fixed length $w$.  With the same notation as before, the estimator of the CJ probability between stock $l$ and $k$ is given by
\begin{equation}
  \hat{p}_w^\text{CJ} = \frac{\sum_{i = 1}^{\left \lfloor \frac{N}{w} \right \rfloor} \mathbf{1}_{ s_{i}^l \geq 1}
  \mathbf{1}_{ s_{i}^k \geq 1}}{\left \lfloor \frac{N}{w} \right \rfloor} \,.
  \label{eq:pcw}
\end{equation}

Our statistical procedure consists in estimating these quantities in real data and compare them, at each time scale, with the 99\% and 95\% confidence bands of the tested model obtained analytically (when possible) or via Monte Carlo simulations.  It should be noted that, since the number of windows that we are considering is greater than one, we are indeed performing a multiple hypothesis test and we have to correct the significance level accordingly.
Among the possible approaches discussed in literature, we decide to adopt the most conservative one, namely the Bonferroni correction. 
This correction amounts to divide the significance level of the single hypothesis by the total number of tested hypotheses in order to achieve a global significance at least of the pre-fixed level. Therefore when the empirical points of $\hat{p}_w^\text{MJ}$  and $ \hat{p}_w^\text{CJ}$ fall inside the confidence bands of a given model we cannot reject the null hypothesis with the given confidence level.

\subsection{A benchmark case: the Poisson model}

In order to show how our testing procedure works, we consider here an important benchmark case, in which the jumps of each stock arisisisisisisise described by an independent Poisson process.
Under this model, the mean and variance of both previous estimators can be computed. For the MJ estimator we have
\begin{equation*}
  \mathbb{E} [\hat{p}_w^\text{MJ}] = p_{w,\lambda}\,, \qquad \text{Var} [\hat{p}_w^\text{MJ}] = \frac{p_{w,\lambda} - p_{w,\lambda}^2}{\left \lfloor \frac{N}{w} \right \rfloor}\,,
\end{equation*}
where $p_{w,\lambda} = \mathbb{P} (\{ s \geq 2 \}) = 1 - \mathrm{e}^{\lambda w}(1 - \lambda w)$, and 
$\lambda$ is the intensity of the Poisson process.

Analogously for the CJ estimator, we obtain
\begin{equation*}
  \mathbb{E} [\hat{p}_w^\text{CJ}] = q_{w,\lambda_l} q_{w,\lambda_k}\,, \qquad \text{Var} [\hat{p}_w^\text{CJ}] = \frac{q_{w,\lambda_l} q_{w,\lambda_k} - 
  q_{w,\lambda_l}^2 q_{w,\lambda_k}^2}{\left \lfloor \frac{N}{w} \right \rfloor}\,, 
\end{equation*}
where $q_{w,\lambda_i} = \mathbb{P} (\{ s^i \geq 1 \}) = 1 - \mathrm{e}^{\lambda_i w}$.

Since both quantities in Equation~(\ref{eq:psw}) and~(\ref{eq:pcw}) correspond to the sum of a large number of indicator functions, the Central Limit Theorem 
implies that their distribution is well approximated by a Normal law whose $p$-values are readily available, allowing the analytical computation of the confidence bands. As usual, the value of the intensity of the Poisson process given by the maximum likelihood estimator is $\hat\lambda = \# \textrm{jumps} / N$.
\begin{figure}[h]
  \begin{center}
    \includegraphics[scale=1.0]{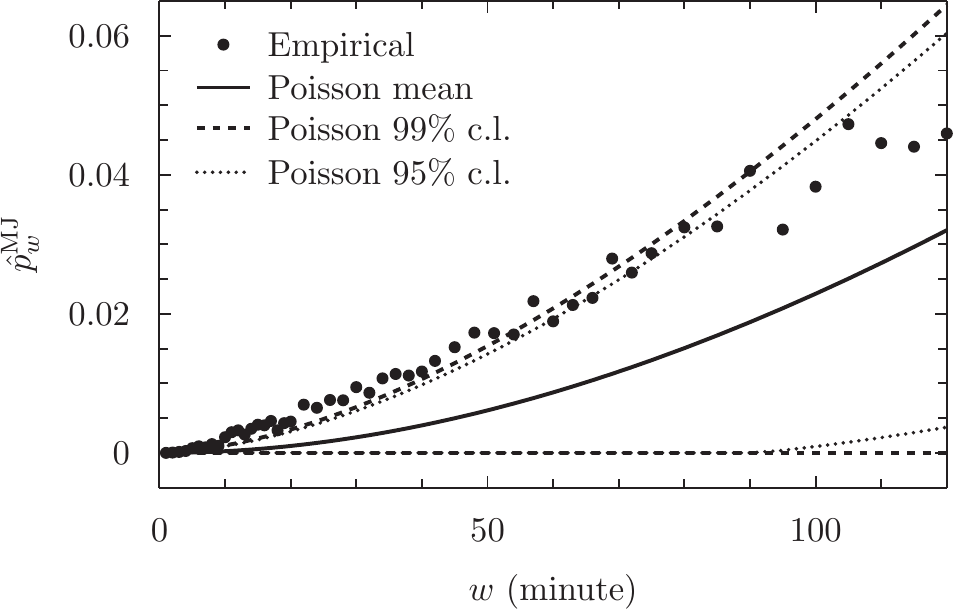}
  \end{center}
  \caption{MJ probability test under Poisson null for the Italian asset Assicurazioni Generali.}
  \label{fig:test-self-poisson}
\end{figure}

In Figure \ref{fig:test-self-poisson} we show the result of our test on the Italian asset Assicurazioni Generali, for which $\lambda=2.4\times 10^{-3}~\text{min}^{-1}$. 
The filled circles correspond to the empirical values of the estimator, the solid line to the theoretical mean, while the dashed and dotted lines establish
the boundaries of the 99\% and 95\% confidence bands, respectively, adjusted with the Bonferroni correction. 
The figure shows that the Poisson model is rejected for both levels. In our data sample this is the typical situation, which occurs for 18 stocks out of 20 (the two exceptions are  represented by the assets Mediobanca and Finmeccanica). We therefore conclude that there is a strong evidence of time 
clustering of jumps and violation of the univariate Poisson model.

\section{Modelling jumps with Hawkes processes}\label{sec:hawkes}

At this stage, the natural step to proceed is to assume a more sophisticated model for the jump process, and to test it as an alternative hypothesis.
The Poisson hypothesis can be relaxed in several respects: among the possible alternatives we can weaken the assumption of identically and independently distributed waiting times,
for example modelling them either as Markov processes or as realizations of a time inhomogeneous Poisson process. 
For instance the extension of the Poisson process that we consider in this paper is the class of point processes known has \textit{Hawkes processes}, 
where the intensity is itself stochastic and tends to increase when a new jump arrives.

\subsection{Univariate case}

In this section we provide the main results needed for the remaining of the paper, but for a complete mathematical treatment of Hawkes and more general point processes
we refer the reader to the comprehensive textbook~\cite{Daley_Vere-Jones:2003}.

A univariate point process $N(t)$\hspace{2pt}\footnote{If $\{t_i\}_{i=1,\ldots,n}$ represents the random sequence of increasing event times $0<t_1<\ldots<t_n$ associated with
the point process, then $N(t) \doteq \sum_{i\geq 1} \mathbf{1}_{t_i \leq t}$ defines the right continuous counting function. In what follows we will refer equivalently to the
process and its counting function.}
is called a Hawkes process if it is a \textit{linear self-exciting process}, defined by the intensity
\begin{equation}
  I(t) = \lambda(t) + \int_{-\infty}^t \nu(t- u) \mathrm{d}N(u)=\lambda(t)+\sum_{t_i<t} \nu(t - t_i)\,,
  \label{eq:hawkes-intensity}
\end{equation}
where $\lambda$ is a deterministic function called the base intensity, $\nu$ is a positive decreasing weight function, and $t_i$ are the jumping times.
The most common parametrization of $\nu$ is given by $\nu(t)=\sum_{j=1}^{P} \alpha_j \mathrm{e}^{-\beta_j t}$, for $t>0$, where $\alpha_j\geq0$
are scale parameters, $\beta_j>0$ control the strength of decay, and the positive integer $P$ is the order of the process.
A particular advantage of the linear Hawkes process of order $P=1$ is that the log-likelihood function can be computed as 
\begin{equation*}
  \mathcal{L}(t_1,\ldots,t_n) = (1-\lambda)t_n -\frac{\alpha}{\beta}\sum_{i=1}^n\left(1-\mathrm{e}^{-\beta(t_n-t_i)}\right)
  + \sum_{i=1}^n \ln (\lambda + \alpha R_i)\,,
\end{equation*}
where the $R$ function satisfies the recursion $R_i = \mathrm{e}^{-\beta(t_i-t_{i-1})}(1 + R_{i-1})$ for $i\geq 2$ and $R_1 = 0$.
In Ref.~\cite{Hawkes:1971} it is shown that the stationarity of the process is guaranteed when $\int_0^\infty \nu(s) \mathrm{d}s < 1$, which in our case reduces to the 
requirement $\alpha/\beta < 1$. If stationarity holds and under the further constraint that the base intensity is constant, the expected number of jumps in an arbitrary 
time interval of length $T$ is given by $\lambda T / (1 - \alpha/\beta)$. The latter observation will be useful for the calibration of the factor models that we will discuss in Section~\ref{sec:factor-model}. The characterization of the asymptotic properties of the the maximum likelihood estimator of the Hawkes process parameters that we employ in this paper has been provided in~\cite{Ogata:1978} and~\cite{Ozaki:1979}, while the simulation algorithm we use is based on the procedure discussed in~\cite{Ogata:1981}, which directly derives from the Shedler-Lewis thinning algorithm,~\cite{Lewis_Shedler:1976}.
\begin{figure}[h]
  \begin{center}
    \includegraphics[scale=0.8]{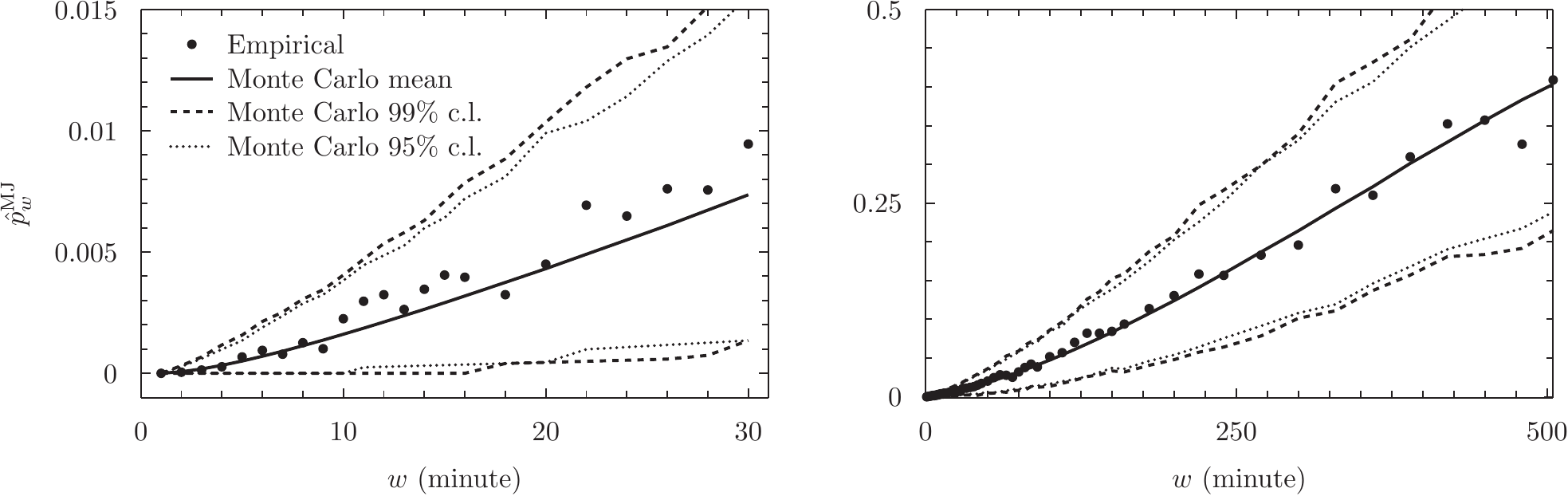}
  \end{center}
  \caption{MJ probability test under Hawkes null for the assets Assicurazioni Generali (left) and Intesa Sanpaolo (right).}
  \label{fig:test-self-hawkes}
\end{figure}

We estimate the Hawkes processes on the univariate series of jumps of the investigated stocks. In Table~\ref{tab:one-dim-hawkes} we report the parameters values with the associated errors and significance values for Generali and Intesa Sanpaolo. All the values are statistically significant. 
\begin{table}
  \begin{center}
    \begin{tabular}{c|l|l}
      \rule[-7pt]{0pt}{19pt}  & Assicurazioni Generali & Intesa Sanpaolo\\
      \hline
      &&\\
      $\lambda~(\text{min}^{-1})$ & $(2.1 \pm 0.2)\times 10^{-3}$ $^{***}$ & $(2.5 \pm 0.2)\times 10^{-3}$ $^{***}$ \\
      $\alpha ~(\text{min}^{-1})$ & $(3.1 \pm 1.3)\times 10^{-2}$ $^{*}  $ & $(5.9 \pm 2.1)\times 10^{-2}$ $^{**} $ \\
      $\beta  ~(\text{min}^{-1})$ & $(2.5 \pm 0.9)\times 10^{-1}$ $^{**} $ & $(4.3 \pm 1.2)\times 10^{-1}$ $^{***}$ \\
      &&\\
      \hline
    \end{tabular}
  \end{center}
  \caption{List of the parameters of the one dimensional Hawkes processes. 
    Significance codes: $\pval < 0.001$ `$^{***}$', $0.001\leq\pval<0.01$ `$^{**}$', $0.01\leq\pval<0.05$ `$^{*}$', and $\pval\geq 0.05$ `~'.}
    \label{tab:one-dim-hawkes}
\end{table}
We then test the Hawkes model in its ability of reproducing the MJ probability at different time scales. The left and right plots in Figure~\ref{fig:test-self-hawkes} are obtained after the calibration of a one dimensional Hawkes process on the jump time series of the assets Assicurazioni Generali and Intesa Sanpaolo, respectively. The figure shows that we can not reject the null both at $1\%$ and at $5\%$ significance  levels (obtained from $N_{MC} = 10^4$ Monte Carlo simulations). 
The result holds both for short window lengths and for longer horizons, an example of the former case is given by Generali for time windows ranging from one minute up to half an hour, while for the latter case we consider Intesa Sanpaolo with horizons running up to one day. In the sections which follow we present the result of the statistical tests only for short horizons, since they usually correspond to the time scales where the most interesting effects take place. However, an analysis extended to longer horizons would have been equally effective.   

We therefore conclude that univariate Hawkes processes are able to capture the empirical dynamics and time clustering of jumps for individual stocks.

\subsection{Bivariate case}

\begin{figure}[t]
  \begin{center}
    \includegraphics[scale=1.0]{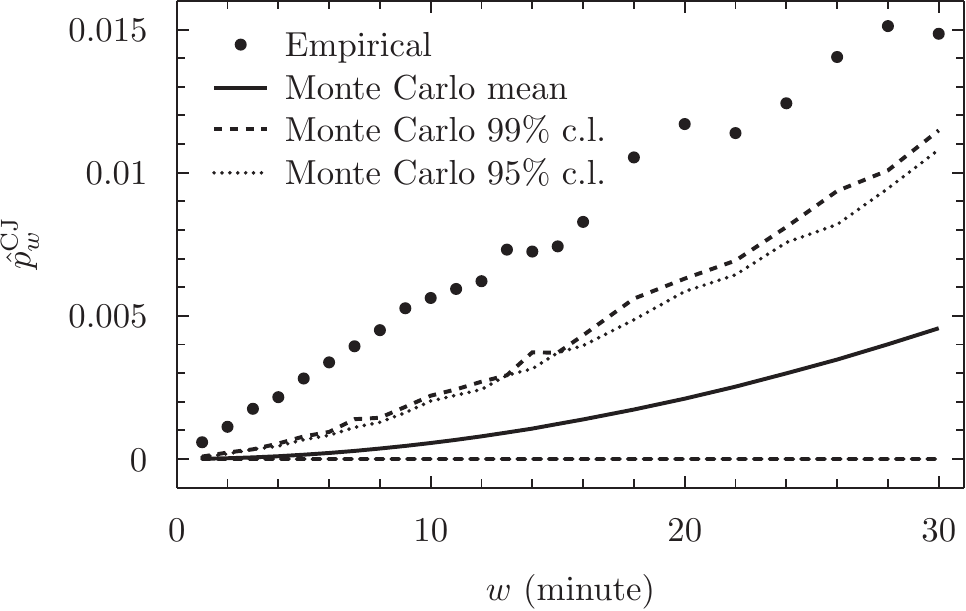}
  \end{center}
  \caption{CJ probability test under independent Hawkes null for the pair of Italian assets Generali - Intesa Sanpaolo.}
  \label{fig:test-cross-hawkes-1d}
\end{figure}

Are independent Hawkes processes able to describe the empirical behaviour of CJ probabilities?
In order to answer this question, we compute the estimator~(\ref{eq:pcw}) on each pair of assets and in Figure~\ref{fig:test-cross-hawkes-1d} we show an example of the results from the test computed on the pair Generali - Intesa Sanpaolo. 
The figure shows that the independent Hawkes process miserably fails in describing the CJ probability at all time scales. This is clearly due, at least in part, to a lack of coupling between the two processes. In order to capture such a dependence we calibrate a bivariate Hawkes process. 

A $K$-dimensional Hawkes process is a linear self-exciting process defined by the multivariate intensity $\boldsymbol{I}(t) = \left(I^1(t),\ldots,I^K(t)\right)'$,
where the $k$-type intensity with an exponential kernel of order one is given by
\begin{equation*}
  I^k(t) = \lambda^k(t) + \sum_{m=1}^K \sum_{t_i^m<t} \alpha_{km}\mathrm{e}^{-\beta_{km}(t - t_i^m)}\,.
\end{equation*}
All the parameters which appear in the above expression are strictly positive, and the stationarity of the process is guaranteed if the spectral radius of the matrix 
$\Gamma = \left(\frac{\alpha_{km}}{\beta_{km}}\right)_{k,m=1,\ldots,K}$ is strictly smaller than one. The parameters $\alpha_{kk}$ and $\beta_{kk}$ are responsible for the self-exciting property of the point process, while the remaining $2K(K-1)$ $\alpha_{km}$ and $\beta_{km}$ capture the cross exciting effect of a jump in the stock $m$ on the process of the asset $k$. When $K=2$ the number of free parameters is equal to ten, and the maximization of the likelihood becomes less trivial than the univariate case. To maximize the likelihood, we initialize the parameters with the values suggested by the one dimensional calibration, and we constrain their value to remain strictly positive. We preliminarily perform $N_{Ann} = 100$ searches of the maximum with the simulated annealing algorithm, then, with the optimal candidate supplied by the stochastic search, we initialize the deterministic search via conjugate gradient. 

\begin{table}
  \begin{center}
    \begin{tabular}{r|l|r|l}
      \hline
      &&&\\
      $\lambda^1  ~(\text{min}^{-1})$ & $(2.0 \pm 0.1)\times 10^{-3}$ $^{***}$ & $\lambda^2  ~(\text{min}^{-1})$ & $(2.3 \pm 0.2)\times 10^{-3}$ $^{***}$ \\
      $\alpha_{11}~(\text{min}^{-1})$ & $(1.6 \pm 0.9)\times 10^{-2}$ $^{*}  $ & $\alpha_{21}~(\text{min}^{-1})$ & $(4.5 \pm 0.1)\times 10^{-4}$ $^{***}$ \\
      $\alpha_{12}~(\text{min}^{-1})$ & $(14~ \pm 0.1)\times 10^{-4}$ $^{***}$ & $\alpha_{22}~(\text{min}^{-1})$ & $(3.4 \pm 1.3)\times 10^{-2}$ $^{**} $ \\
      $\beta_{11} ~(\text{min}^{-1})$ & $(3.7 \pm 1.6)\times 10^{-1}$ $^{*}  $ & $\beta_{21} ~(\text{min}^{-1})$ & $(7.0 \pm 18~)\times 10^{-1}$ $^{~}  $ \\
      $\beta_{12} ~(\text{min}^{-1})$ & $(3.2 \pm 3.4)\times 10^{-1}$ $^{~}  $ & $\beta_{22} ~(\text{min}^{-1})$ & $(4.9 \pm 1.6)\times 10^{-1}$ $^{***}$ \\
      &&&\\
      \hline
    \end{tabular}
  \end{center}
  \caption{List of the parameters of the bivariate Hawkes process for Generali and Intesa Sanpaolo (labelled with the indices 1 and 2, respectively). 
  Significance codes: $\pval < 0.001$ `$^{***}$', $0.001\leq\pval<0.01$ `$^{**}$', $0.01\leq\pval<0.05$ `$^{*}$', and $\pval\geq 0.05$ `~'.}
  \label{tab:two-dim-hawkes}
\end{table}
\begin{figure}[h]
  \begin{center}
    \includegraphics[scale=1.0]{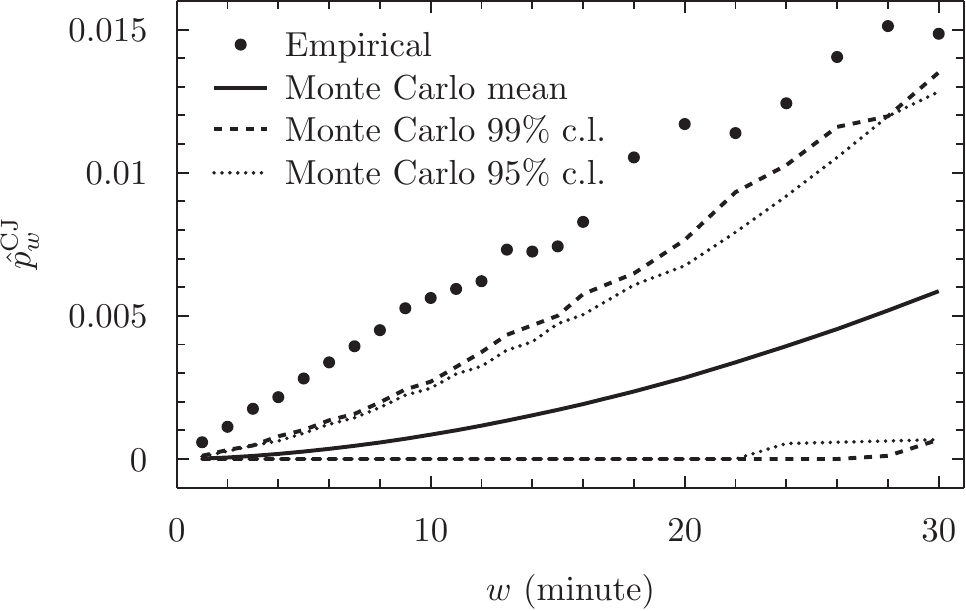}
  \end{center}
  \caption{CJ probability test under bivariate Hawkes null for Generali - Intesa Sanpaolo.}
  \label{fig:test-hawkes-2d}
\end{figure}

In Table~\ref{tab:two-dim-hawkes} we present the results of the calibration of a bi-dimensional Hawkes on the joint jumps process
of Generali-Intesa. We then use Monte Carlo simulations to compute confidence bands for CJ probability as shown in Figure~\ref{fig:test-hawkes-2d}. 
It is clear from the figure that  bivariate exponential Hawkes fails to describe the cross sectional 
clustering of jumps observed in real data. This might be due to several facts: the value of the cross exciting constants $\alpha_{12}$ and $\alpha_{21}$
is at least of one order of magnitude smaller than that of the constants $\alpha_{11}$ and $\alpha_{22}$. Moreover, the time constants $\beta_{12}$ and 
$\beta_{21}$ associated to the cross excitations are poorly statistically significant. 
One possible reason could be the non reliability of the parameters since the optimization procedure is performed in a high dimensional space on a strongly non linear objective function.
However, in this respect, we perform numerous experiments in a Monte Carlo framework. One test is the simulation of a bivariate Hawkes process, whose parameters 
are chosen equal to the values of Table~\ref{tab:two-dim-hawkes}. On each random realization, we apply the above optimization scheme and record the optimal values.
With a statistics of $10^{3}$ values per parameter, we measure no significant bias induced by the multidimensional maximization procedure. Moreover,
we simulate $10^{3}$ independent pairs of univariate Hawkes of length 44440 minutes, whose parameters are given in Table~\ref{tab:one-dim-hawkes}.
We then perform the ten dimensional optimization on each copy, and find a confirmation of the tendency of the cross-exciting constants to significantly decrease.
In light of these numerical evidences, we tend to marginalize the role played by the optimization scheme. 
Finally, we believe that the main reason why Hawkes processes fail to describe the cross-sectional dependence is that they are not 
designed by construction to capture synchronous effects, which in the case under our consideration seem to dominate the dependence between jumps.

\section{A factor model approach to systemic jumps}\label{sec:factor-model}

In this section we abandon the idea of an $N$ dimensional point process, and we develop a different approach, which we will refer to as a jump factor model. 
The intuition which drives the modelling is that a stock jumps both because it is triggered by jumps of a market common factor and because of an 
idiosyncratic term.

The first issue that we want to clarify is if, even in a Poisson framework, the factor mechanism is able to describe the empirical dependence structure. Then we will propose a 
scalable model, which ideally should be effective and also sufficiently simple and robust.

\subsection{Bivariate Poisson factor model}

In order to illustrate the main idea, we start with a toy model which enlightens the role of the market
factor and clarifies the mechanism which generates the dependence structure. However, this is
an unrealistic model, where the behaviour of the assets is completely determined by the evolution of the factor process.

We assume that there is one unobserved market factor point process. When the factor jumps, the stock $S_1$ jumps with probability $p_1$ and the stock 
$S_2$ jumps with probability $p_2$. If the factor does not jump, neither the first nor the second stock can jump. Obviously the converse is not true.
If the market factor is described by a Poisson process of intensity $\lambda_F$, then the expected number of factor's jumps over the period $T$ is 
given by $\lambda_F T$.
In order to fix the free parameters of the model, $\lambda_F$, $p_1$, and $p_2$, we use the following  relations
\begin{eqnarray}
  p_1 ~ \lambda_F T &= n_1\,,\nonumber\\
  p_2 ~ \lambda_F T &= n_2\,,\nonumber\\
  p_1 ~ p_2 ~ \lambda_F T &= n_{12}\,,\label{eq:p1-p2}
\end{eqnarray}
where $n_1$ and $n_2$ are the observed number of jumps of the stock $S_1$ and $S_2$, respectively, while $n_{12}$ 
represents the observed number of cojumps among $S_1$ and $S_2$ within the one minute sampling interval.
The first two equations thus require that the expected number of jumps of the single assets matches the realized values. 
The latter relation guarantees that, on average, the realized number of cojumps among the assets corresponds to the theoretical expectation.
The system of Equations~(\ref{eq:p1-p2}) can be inverted and provides a direct way to express the parameters of 
the model in terms of the observable quantities $T$, $n_1$, $n_2$, and $n_{12}$:
\begin{equation*}
  \lambda_F = \frac{n_1 n_2}{n_{12}}\frac{1}{T}\,, \quad p_1 = \frac{n_{12}}{n_2}\,,\quad\text{and}\quad  p_2 = \frac{n_{12}}{n_1}\,.
\end{equation*}

\begin{figure}[h]
  \begin{center}
    \includegraphics[scale=1.0]{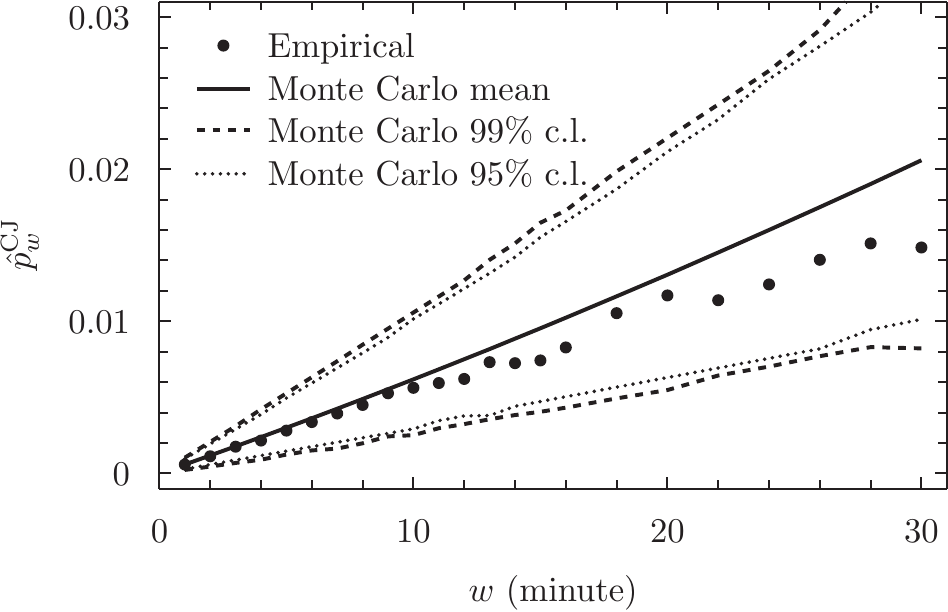}
  \end{center}
  \caption{CJ probability test under Poisson factor null for Generali - Intesa Sanpaolo.}
  \label{fig:test-poisson-factor}
\end{figure}

In Figure~\ref{fig:test-poisson-factor} we show the result of the test of the cojumps estimator $\hat{p}_w^\mathrm{CJ}$ against a null represented by the Poisson factor model
that we have just described.
The time period that we consider corresponds to the usual interval of 88 days, during which Generali and Intesa Sanpaolo jump 103 and 127 times, respectively, 
while the realized number of cojumps $n_{12}$ is equal to 26. 
From these values we obtain a Poisson intensity equal to $\lambda_F=1.1\times 10^{-2}$ per minute, and probabilities $p_1=0.20$ and $p_2=0.25$. 
The confidence intervals for the null are estimated drawing $N_{MC}=10^4$ paths from the Poisson factor process, 
and then thinning each realization with Bernoulli variables of probability $p_1$ and $p_2$. 

The results provided by this simple model are very satisfactory and should convince the reader that the proposed mechanism is adequate to capture the cross dependence between jumps. 
However, the model is quite unrealistic in several respects: (i) the assets can not jump independently of the factor, (ii) the Poisson nature of the process leads to a severe 
underestimation of the realized number of multiple jumps of the same stock, and (iii) the approach has not a straightforward extension to an arbitrary number of assets. 
In order to amend all these drawbacks, in the next section we will discuss a more general and flexible model, rooted on the factor idea, 
but able to achieve a higher level of realism and scalability.

\subsection{$N$ dimensional Hawkes factor model with idiosyncratic components}\label{sec:n-factor}

We now consider a set of $N$ assets. With respect to the previous model, we need to introduce explicitly a proxy of the common factor process. 
The naif idea of a proxy of the factor based on those events when a cojump among all the $N$ stocks happens realistically does not work. 
In the data sample of twenty stocks that we consider in the current study, we experience just a single event when all the stocks jump simultaneously. 
The proxy that we propose is characterized by a counting function which increases by one unit whenever we detect a cojump 
which involves a group of assets whose number $J$ conflicts with the null of independence. We therefore need to identify the threshold value $\bar{J}$ compatible 
with independence. In general it will be time dependent, $\bar{J}_t$, and will vary between one and $N$. 
In order to detect $\bar{J}$ we propose a rigorous statistical methodology.
However, we preliminarily want to convince the reader that the range of variability of $\bar{J}$ for the case under study $N=20$ is quite narrow.
First of all, we can exclude the case $\bar{J}=1$, since otherwise we would consider every jump that occurs on the market as potentially systemic.
On the other side, if we fix $\bar{J}$, by definition then we would consider the events which involve $J<\bar{J}$ jumps as likely occurrences. 
As shown by the bootstrap experiment in Section~\ref{sec:systemic_jumps}, the probability associated with events involving more than two stocks is extremely low,
and this would fix $\bar{J}=3$. However, the null represented by the bootstrapped data is too extreme, since it does not take into account any effect of self excitation. 
Heuristically, we conclude that $\bar{J}\simeq 4$.

We now present a rigorous procedure that supports the above heuristic argument.
To each stock in our sample we associate a finite number of event times $\{t_i^s\}$, where $s=1,\ldots,N$ and $i=1,\ldots,n_s$ label the assets and the asset idiosyncratic jumps, respectively. In equivalent terms, each asset is characterized by a counting function $N^s(t)$, and $n_s=N^s(T)$ is the observed number of jumps of the stock $s$. According to the procedure of jumps' identification, the $t_i^s$'s are measured in minutes and take only integer values running from one to $T$. A counting process $N^s(t)$ is associated with an intensity function $I^s(t)$ defined by
\begin{equation*}
  \mathbb{P}\left(N^s(t) \text{ has a jump in } [t,t+\Delta t] | \mathcal{F}_t \right) = I^s(t) \Delta t\,,
\end{equation*}
for $\Delta t \rightarrow 0^+$, where $\mathbb{P}$ stands for the probability and $\mathcal{F}_t$ corresponds to the history of the process up to present time $t$.
We then assume that the counting processes which describe our assets correspond to $N$ independent univariate Hawkes processes and we estimate the parameters
which characterize the intensities $I^s(t)$ for $s=1,\ldots,N$ via maximum likelihood. We fix $\Delta t$ equal to one minute and via Equation~(\ref{eq:hawkes-intensity}), for each $t=1,\ldots,T$, we can compute the vector of probabilities $\boldsymbol{\pi}_t=\left(\pi_t^1,\ldots \pi_t^N\right)'=\left(I_t^1 \Delta t,\ldots, I_t^N\Delta t\right)'$.
We test if the number of jumps that we observe at time $t$ is compatible with the cross independence among the processes. Under the null hypothesis, 
the discrete probability of the event $J_t=j$ reads 
\begin{equation*}
  \mathbb{P}\left(J_t=j\right) = \sum_{1\leq l_1<\ldots<l_j\leq N} \pi_t^{l_1}\ldots \pi_t^{l_j} 
  \prod_{k \in \{1,\ldots,N\} \setminus \{l_1,\ldots,l_j\}}\left(1-\pi_t^{k}\right)\,.
\end{equation*}
Since we repeat the test $T$ times, we adjust the significance level with the Bonferroni's correction. If at time $t^F$ we reject the null, 
we attribute the event to a systemic shock, and we remove it from the set $\{t_i^s\}$. The procedure has to be iterated as many times as required in order 
to remove all the systemic jumps. 

The advantages of the above procedure are manifold. Specifically, the set of events $\{t_i^F\}$ with $i=1,\ldots,n_F$ identifies the $n_F$ jumps of the common factor, but not less important each reduced set $\{t_{i^\prime}^s\}_{i^\prime = 1,\ldots,n_{s}^\prime}=\{t_{i}^s\}_{i = 1,\ldots,n_{s}}\setminus\{t_{j}^F\}_{j = 1,\ldots,n_{F}}\subseteq \{t_i^s\}_{i = 1,\ldots,n_{s}}$ for $s=1,\ldots,N$ corresponds in a natural way to the set of the $n_{s}^\prime$ jumps of the $s$-th idiosyncratic component. The latter result represents a second major improvement with respect to the bivariate model. Moreover, the entire procedure is easy scalable, since it does not depend critically on the dimension of the portfolio. Last but not least, we are not a priori fixing the nature of the point processes which describe the factor and the idiosyncratic components. We can apply the log-likelihood approach described in~\ref{sec:hawkes} to the univariate sequence of the event times, fix the value of the parameters of the factor, $\lambda_F$, $\alpha_F$ and $\beta_F$, and of the idiosyncratic components, $\lambda_s$, $\alpha_s$ and $\beta_s$, for $s=1,\ldots,N$, estimate the associated $p$-values, and, eventually, if it is the case, reject the Hawkes process in favour of the Poisson description.

To sum up, the multivariate model describes the extreme events which occur in a portfolio of stocks as a superposition of systemic shocks, propagating to the
$s$-th asset with probability $p_s$, and jumps specific of the single assets. 
In order to estimate the vector of probabilities $\boldsymbol{p}=\left(p_1,\ldots,p_N\right)'$, we 
replace the system of equations given by~(\ref{eq:p1-p2}), with the new relations
\begin{eqnarray*}
  p_1 ~ \frac{\lambda_F}{1-\alpha_F/\beta_F} T &=& n_{1} - n_{1}^\prime\,,\\
  &\vdots&\\
  p_N ~ \frac{\lambda_F}{1-\alpha_F/\beta_F} T &=& n_{N} - n_{N}^\prime\,,
\end{eqnarray*}
where $\lambda^F T /(1-\alpha_F / \beta_F)$ corresponds the average number of factor's jumps. 
The above equations force the expected number of cojumps among the factor and the $s$-th stock to balance the realized number of shocks. 
\begin{figure}[h]
  \begin{center}
    \includegraphics[scale=0.8]{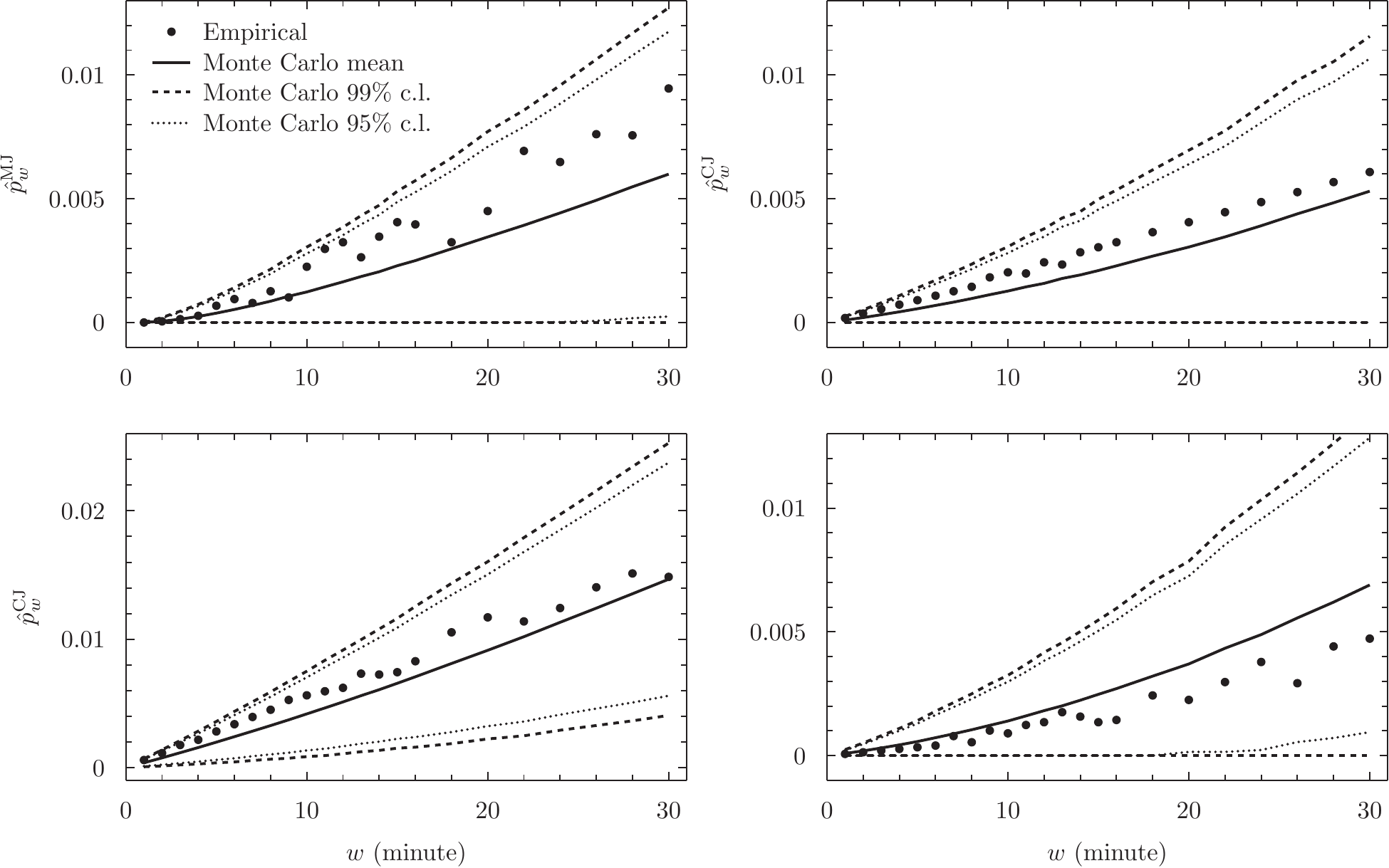}
  \end{center}
  \caption{From the top left clockwise: MJ probability test under $N$ factor model null for the asset Generali;
  CJ probability test for the pairs Generali-Mediobanca, Generali-Banca Popolare Milano, and Generali-Intesa Sanpaolo.}
  \label{fig:test-hawkes-hawkes}
\end{figure}

We test our model on the data set of twenty stocks analysed in the first part of this paper, and we show in Figure~\ref{fig:test-hawkes-hawkes} the results 
of the tests performed over the assets Generali, Mediobanca, Banca Popolare di Milano, and Intesa Sanpaolo.
The plots clearly show the ability to capture both the self and the cross dependence among jumps,
which is a remarkable feature of a factor model easy to implement and calibrate on the data. Moreover, by construction, the model is genuinely scalable, and this fact 
makes it a viable alternative to more sophisticated but computationally complex models. In Table~\ref{tab:factor} we present the value of the parameters of the 
common factor, with the associated standard errors and $p$-values. 
From the measured confidence for the scale and decay length we can reject the null hypothesis of a Poisson model for the factor. 
In Table~\ref{tab:components} we report the remaining parameters for the idiosyncratic components.

It is important to notice that we have performed a Monte Carlo experiment by drawing  $N_{MC}=10^3$ scenarios with the values given in Tables~\ref{tab:factor} and~\ref{tab:components} and we have then  
re-estimated the parameters on each copy. This experiment has shown that the entire procedure is quite robust, but the comparison with the process $N_t^F$, 
which is known in the artificial Monte Carlo framework, evidences a systematic underestimation of the number of jumps of the common factor. 
The mis-identification is due to the fact that even when the factor jumps, there is always a small, but finite, probability that only a small number of assets 
below the detectability threshold jumps too. In this case, our methodology does not detect a systemic event. The mis-identification probability depends on the true values of 
$(p_1,\ldots,p_N)$, and for those fixed in the Monte Carlo experiment, it amounts approximately to $7.5\%$.
Nonetheless, the bias reduces when the number of assets increases, and tends to zero for $N\rightarrow \infty$.
\begin{table}
  \begin{center}
    \begin{tabular}{l|l}
      \rule[-7pt]{0pt}{19pt}  & Hawkes factor\\
      \hline
      &\\
      $\lambda_F $ & $(2.0 \pm 0.2)\times 10^{-3}$, $p$-value: $<$ 0.001\\
      $\alpha_F  $ & $(4.9 \pm 1.9)\times 10^{-2}$, $p$-value: 0.0105\\
      $\beta_F   $ & $(3.3 \pm 1.1)\times 10^{-1}$, $p$-value: 0.0021\\
      &\\
      \hline
    \end{tabular}
  \end{center}
  \caption{List of the parameters of the common factor process.}
  \label{tab:factor}
\end{table}
\begin{sidewaystable}
  \begin{center}
    \begin{tabular}{l|l|l|l|l}
      \rule[-7pt]{0pt}{19pt} ISIN & $p_{s}$ & $\lambda_{s}~(\times 10^{-3})$, $p$-value & $\alpha_s~(\times 10^{-2})$, $p$-value & $\beta_s~(\times 10^{-1})$, $p$-value \\
      \hline
      &&&&\\
      IT0000062072 & 0.31 & $1.4 \pm 0.2$, $^{(*)}$ & $2.6 \pm 1.4$, 0.067 & $2.2 \pm 1.0$, 0.030 \\
      IT0000062957 & 0.12 & $1.1 \pm 0.2$, $^{(*)}$ & $1.1 \pm 1.3$, 0.401 & $2.3 \pm 2.6$, 0.386 \\
      IT0000064482 & 0.10 & $2.2 \pm 0.2$, $^{(*)}$ & $3.7 \pm 1.6$, 0.020 & $2.6 \pm 1.1$, 0.014 \\
      IT0000068525 & 0.24 & $1.2 \pm 0.2$, $^{(*)}$ & $0.9 \pm 0.4$, 0.020 & $0.3 \pm 0.1$, 0.021 \\
      IT0000072618 & 0.48 & $1.6 \pm 0.2$, $^{(*)}$ & $0.6 \pm 0.6$, 0.277 & $1.0 \pm 1.1$, 0.365 \\
      IT0001063210 & 0.08 & $1.0 \pm 0.2$, $^{(*)}$ & $0.6 \pm 0.5$, 0.174 & $0.6 \pm 0.4$, 0.125 \\
      IT0001334587 & 0.17 & $2.9 \pm 0.3$, $^{(*)}$ & $6.2 \pm 1.8$, 0.001 & $3.0 \pm 0.8$,  $^{(*)}$\\
      IT0001976403 & 0.31 & $1.7 \pm 0.2$, $^{(*)}$ & $2.8 \pm 1.2$, 0.015 & $1.7 \pm 0.6$, 0.006 \\
      IT0003128367 & 0.39 & $2.6 \pm 0.3$, $^{(*)}$ & $2.4 \pm 0.8$, 0.004 & $1.2 \pm 0.4$, 0.002 \\
      IT0003132476 & 0.37 & $2.2 \pm 0.2$, $^{(*)}$ & $1.3 \pm 0.5$, 0.016 & $0.9 \pm 0.3$, 0.012 \\
      IT0003487029 & 0.18 & $1.0 \pm 0.2$, $^{(*)}$ & $5.1 \pm 2.8$, 0.069 & $3.9 \pm 1.7$, 0.019 \\
      IT0003497168 & 0.31 & $1.8 \pm 0.2$, $^{(*)}$ & $0.7 \pm 0.4$, 0.126 & $0.4 \pm 0.3$, 0.232 \\
      IT0003856405 & 0.11 & $1.6 \pm 0.2$, $^{(*)}$ & $0.8 \pm 0.4$, 0.087 & $0.7 \pm 0.3$, 0.058 \\
      IT0004176001 & 0.14 & $0.9 \pm 0.2$, $^{(*)}$ & $0.7 \pm 0.3$, 0.024 & $0.2 \pm 0.1$, 0.018 \\
      IT0004231566 & 0.16 & $1.7 \pm 0.2$, $^{(*)}$ & $1.8 \pm 0.9$, 0.056 & $1.5 \pm 0.7$, 0.031 \\
      IT0004623051 & 0.15 & $1.8 \pm 0.2$, $^{(*)}$ & $2.4 \pm 0.9$, 0.008 & $1.0 \pm 0.4$, 0.006 \\
      IT0004644743 & 0.21 & $1.4 \pm 0.2$, $^{(*)}$ & $3.6 \pm 1.5$, 0.014 & $1.9 \pm 0.7$, 0.003 \\
      IT0004781412 & 0.37 & $1.4 \pm 0.2$, $^{(*)}$ & $1.0 \pm 0.4$, 0.025 & $0.5 \pm 0.2$, 0.007 \\
      LU0156801721 & 0.17 & $0.8 \pm 0.1$, $^{(*)}$ & $1.0 \pm 0.7$, 0.147 & $0.7 \pm 0.4$, 0.092 \\
      NL0000226223 & 0.22 & $1.0 \pm 0.2$, $^{(*)}$ & $3.2 \pm 1.6$, 0.048 & $1.2 \pm 0.7$, 0.073 \\
      &&&&\\
      \hline
    \end{tabular}
  \end{center}
  \caption{List of the parameters of the idiosyncratic components; $^{(*)} < 0.001$.}
  \label{tab:components}
\end{sidewaystable}

\section{Conclusions and Perspectives}

The detection techniques that we develop in Sections~\ref{sec:data_handling} and~\ref{sec:jump_identification} show that a large number of jumps is present 
in financial time series. Even though the identification process of the extreme events suffers from some dependence on the details of the detection method, we believe
that the idea of intersecting the different methodologies partially amends this drawback.
Relying on the correct identification of jumps, we find that,
as far as individual stocks are concerned, jumps are clearly not described by a Poisson process.
The evidence of time clustering can be accounted for and modelled by means of linear self-exciting Hawkes processes.
Moving to a cross-sectional perspective, we identify a significant number of systemic events, especially simultaneous cross jumps,
that can not be reduced to a purely random effect.
We have provided quantitative arguments against the idea of modelling this effect in terms of multidimensional Hawkes processes.
The simultaneity of events is not captured by this class of point processes, and the increase of dimensionality of the parameters
space associated with the multivariate model is discouraging.
In Section~\ref{sec:n-factor} we propose a one factor model which is able to describe the main features that characterize the departure from a random behaviour of jumps,
namely, the time clustering of jumps on individual stocks, the large number of simultaneous systemic jumps, and the time lagged cross excitation between different stocks.

This work opens interesting perspectives for future research. It would be of great interest to see whether these results would be confirmed by repeating the analysis on a wider collection of assets. Promising directions could be enlarging the number of investigated stocks, extending the analysis to other markets, or also considering different but related securities, such as equities, futures, options and other derivative products. Another extension of this work could be the study of the properties of jumps for a given group of securities in different periods, to assess for example whether changes in the regulations of a market have an impact on the frequency of systemic events.

Moving from a descriptive point of view to one that investigates the origin of the behaviour of jumps and cojumps, the research direction we consider more promising is the study of the order book in proximity of extreme systemic events, that is, cojumps involving a large number of assets. The perspective to take should be to explore the cross direction at a fixed time, more than the time direction for a single asset. The bid-ask spread, the depth of the book, the asymmetry of buy and sell volumes are all quantities whose dynamics may reveal interesting features related to systemic events.

The high level of synchronization between the jumps of different assets that we empirically observe in our data calls for possible explanations. Even if our paper is mainly methodological, we believe that some comments are needed in order to explain this fact observed in (modern) financial markets. Financial markets are becoming increasingly interconnected at a high speed due to several reasons, first of all the increased automation of the trading process and of the information processing. High frequency trading strategies, statistical arbitrageurs, hedging strategies could be partly responsible of the large number of cojumps we observe in our sample. Certainly, we believe, a proper modelling of the jump process in a systemic context is important for regulators and for investors in order to assess in a reliable way the level of risk of a market or of a large portfolio of assets.    

\section*{Acknowledgements}

GB and LMC acknowledge the support of the Scuola Normale Superiore grant GR12CALCAG ``Entropia di Shannon ed efficienza informativa dei mercati finanziari''. 
FL acknowledges partial	support from grant SNS11LILLB ``Price formation, agents heterogeneity, and market efficiency''. 
FC, FL, and SM acknowledge partial support of the European Union Seventh Framework Programme FP7/2007-2013 under grant agreement CRISIS-ICT-2011-288501.
MT acknowledges Flavio Baronti and Francesco Pantaleo for technical support. 
Finally, the authors wish to thank Enrico Melchioni for the constant encouragement.

\appendix
\section{Data Cleaning}\label{app:data_cleaning}
\subsection{Outlier and split/merge detection}\label{sub:brownlees_gallo}
To detect and remove anomalous values (often referred to as outliers) that may be present in recorded data, we use the cleaning algorithm 
proposed by Brownlees and Gallo in~\cite{Brownlees_Gallo:2006}. The algorithm identifies the price records at tick-by-tick level which are too 
distant from a mean value calculated in their neighbourhood. More precisely, a price observation $p_i$ is regarded as anomalous and removed if
\[ |p_i - \bar{p}_i (k)| \geq c \, s_i (k) + \gamma , \]
where $\bar{p}_i (k)$ and $s_i (k)$ denote respectively the $\delta$-trimmed sample mean and sample standard deviation of the closest $k$ observations 
around $i$, $c$ is a constant which amplifies the standard deviation and $\gamma$ is a granularity parameter useful in the case of $k$ equal prices producing 
a zero variance. We take $k = 60$, $\delta = 10\%$, $c = 3$, $\gamma = 0.05$. 

A \emph{forward stock split} (or, simply, a \emph{stock split}) is an operation consisting in an increase of the number of shares of a company and in a simultaneous
adjustment of the price so that the market capitalization of the company remains the same. A \emph{reverse stock split} (also called a \emph{stock merge}) is 
the opposite operation, leading to a decrease of the number of shares and to a corresponding increase of the price. A split is \emph{$m$-for-$n$} if $m$ new shares
are released for every $n$ old ones, with a simultaneous price change from $p$ to $p \, n / m$. 

\subsection{Volatility auctions} \label{sec:vola_auction}

\begin{figure} 
  \begin{center}
    \includegraphics[scale=0.8]{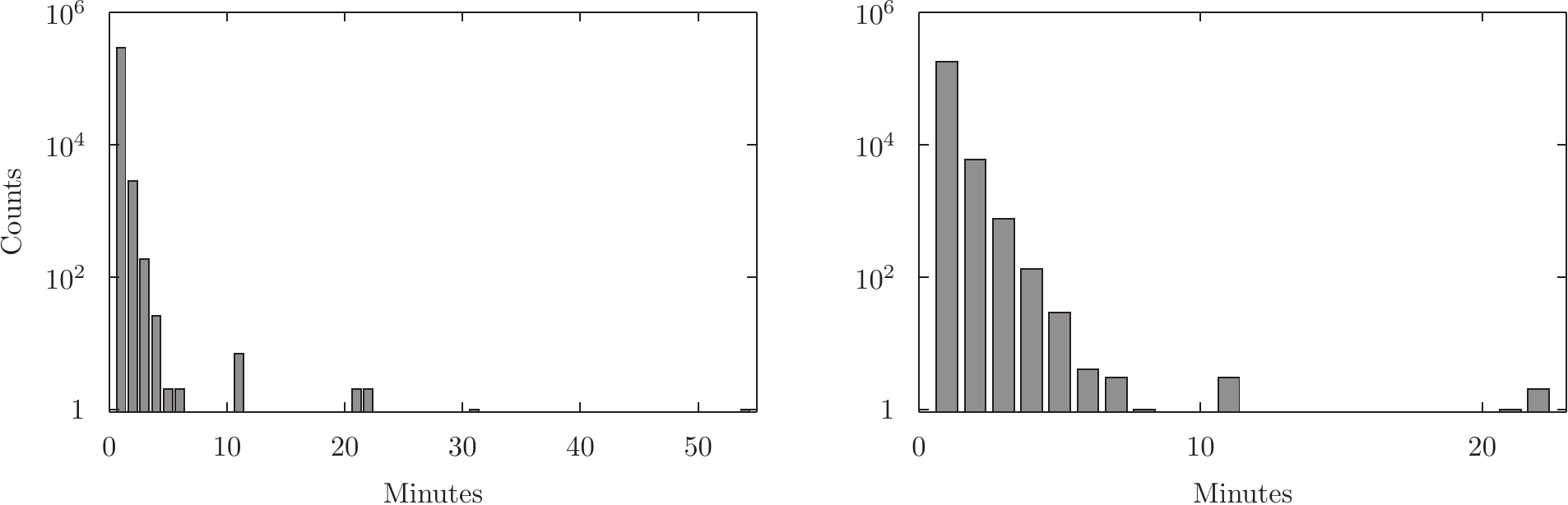}
  \end{center}
  \caption{The distributions of intertrade times for stocks Fiat (left panel) and Telecom Italia (right panel).}
  \label{fig:intertrade_histograms}
\end{figure}

According to the rules stated in \textit{Rules of the markets organised and managed by Borsa Italiana S.p.A.} and 
\textit{Instructions accompanying the Rules of the markets organised and managed by Borsa Italiana S.p.A.} (\cite{Rules:2012,Instructions:2012}), 
whenever a stock's price gets too far from a reference value, Borsa Italiana is obliged to start a volatility auction phase. 
During this phase, which has a duration between 10 and 11 minutes, trades are suspended and a new reference value for the price is sought. 
Possibly, if no valid price is reached during the volatility auction phase, another such phase starts immediately after the first one and so on. 
We treat all intertrade times of at least ten minutes as volatility auction periods (the distributions of intertrade times shown in Figure~\ref{fig:intertrade_histograms} suggest this is correct) and consider as \emph{not available} the returns at the sampling times falling in these periods.
In Figure~\ref{fig:intertrade_histograms} we report the empirical distributions of the intertrade times for the assets Fiat and Telecom Italia. 
In both histograms we can easily recognize the presence of volatility auctions, which manifest in terms of the peaks appearing on the far
right tail in correspondence of intertrade times close to the multiples of eleven minutes.

\subsection{Intraday pattern} \label{sec:intraday_pattern}

We filter out the intraday volatility pattern from returns by means of a simple model with intraday volatility factors. Returns at intraday time $t$ are rescaled by a factor $\zeta_t$, which is calculated as the average, over all days, of adjusted absolute returns at time $t$. More precisely, if $\tilde{r}_{d,t}$ is the raw return of day $d$ and intraday time $t$, we define the rescaled return
\[ r_{d,t} = \frac{\tilde{r}_{d,t}}{\zeta_t} , \]
where
\[ \zeta_t = \frac{1}{N_{\text{days}}} \sum_{d'} \frac{|r_{d',t}|}{s_{d'}} , \]
with $N_{\text{days}}$ indicating the number of days in the sample and $s_{d'}$ the standard deviation of absolute intraday returns of day $d'$.

\begin{figure}
  \centering
  \includegraphics[scale=1.0]{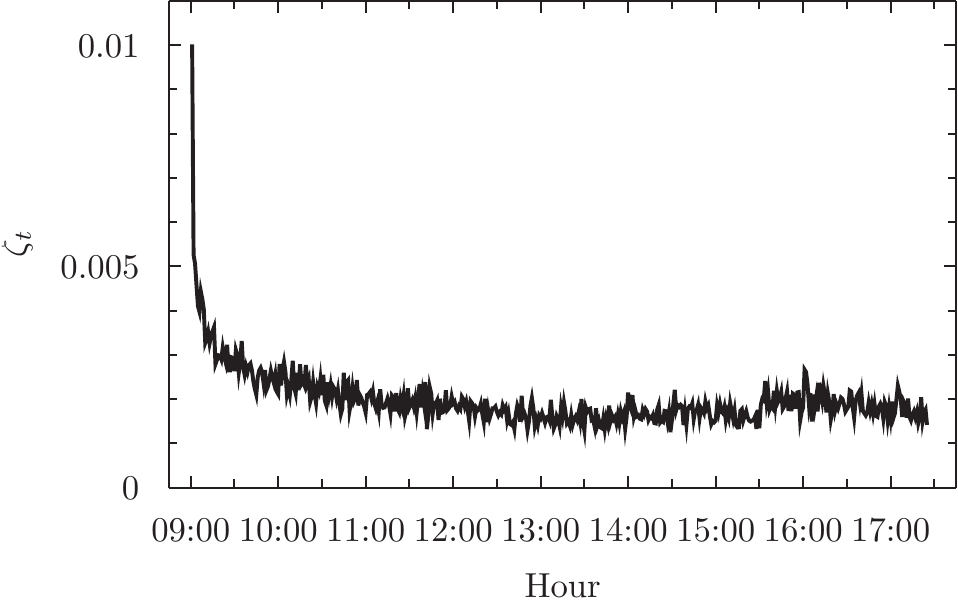}
  \caption{The intraday volatility pattern of Monte dei Paschi di Siena.}
  \label{fig:intra_vol_pat}
\end{figure}

We show in Figure~\ref{fig:intra_vol_pat} the intraday volatility pattern (that is, the factors $\zeta_t$) for Monte dei Paschi di Siena. 
We notice that, although the profile is somewhat noisy because of the relatively small number of days on which the average pattern is calculated, 
a clear change in average intraday volatility is identifiable around time 15:30, that is, when the New York Stock Exchange opens.

\subsection{Volatility proxy} \label{sec:vol_proxy}

For estimating the local volatility, we use the \emph{realized absolute variation} and the 
\emph{realized bipower variation}~\cite{Andersen_Bollerslev:1997,Barndorff-Nielsen_Shephard:2004}. 
Let the logarithmic prices $p (t)$ be generated by a process
\begin{equation} \label{eq:price_process}
  \ud p (t) = \mu (t) \, \ud t + \sigma (t) \, \ud W (t) ,
\end{equation}
with $\mu (t)$ a finite variation process, $\sigma (t)$ a c\`adl\`ag volatility process, $W (t)$ a standard Brownian motion. Let the interval $[0,t]$ be divided into subintervals of the same length $\delta$ and denote by $r_i$ the return $p (i \delta) - p ((i-1) \delta)$. Then the following probability limits hold for the \emph{realized absolute variation} and the \emph{realized bipower variation}:
\begin{equation*}
  \text{p}-\lim_{\delta \searrow 0} \delta^{\frac{1}{2}} \sum_{i=1}^{\lfloor t/\delta \rfloor} |r_i| =
  \mu_1 \int_0^t \sigma (s) \, \ud s ,
\end{equation*}
\begin{equation} \label{eq:bipower_plim}
  \text{p}-\lim_{\delta \searrow 0} \sum_{i=1}^{\lfloor t/\delta \rfloor - 1} |r_i||r_{i+1}| =
  \mu_1^2 \int_0^t \sigma^2 (s) \, \ud s ,
\end{equation}
where $\mu_1 = \text{E} (|u|) = \sqrt{\frac{2}{\pi}} \simeq 0.797885$, $u \sim \mathcal{N} (0,1)$.
The asymptotic result~(\ref{eq:bipower_plim}) continues to hold when a jump component is added to the continuous process~(\ref{eq:price_process}). 
However, for finite $\delta$, price jumps are a source of bias in estimation of volatility through the realized bipower variation. A solution to this 
problem is to use the \emph{threshold bipower variation} (see~\cite{Corsi_etal:2010}), which takes into account only returns smaller than a certain 
threshold. We follow this idea and estimate volatility through returns which are not identified as jumps, that is, returns whose absolute value is not larger than 4 times the local volatility. Using exponentially weighted moving averages instead of flat ones, our estimators are recursively defined by the equations
\begin{equation*}
  \hat{\sigma}_{\text{abs},t} = \mu_1^{-1} \alpha |r_{t^\prime}| + (1 - \alpha) \hat{\sigma}_{\text{abs},t-1} ,
\end{equation*}
where $t^\prime$ is such that $t^\prime \leq t-1$, $\frac{|r_{t^\prime}|}{\hat{\sigma}_{\text{abs},t^\prime}} \leq 4$ and $\frac{|r_\tau|}{\hat{\sigma}_{\text{abs},\tau}} > 4$ for each $t^\prime < \tau < t$, and
\begin{equation*}
  \hat{\sigma}_{\text{bv},t}^2 = \mu_1^{-2} \alpha |r_{t^{\prime\prime}}| |r_{t^\prime}| + (1 - \alpha) \hat{\sigma}_{\text{bv},t-1}^2 ,
\end{equation*}
where $t^{\prime\prime}$ and $t^\prime$ are such that 
$t^{\prime\prime} < t^{\prime} \leq t-1$, $\frac{|r_{t^{\prime\prime}}|}{\hat{\sigma}_{\text{bv},t^{\prime\prime}}} \leq 4$, $\frac{|r_{t^\prime}|}{\hat{\sigma}_{\text{bv},t^\prime}} \leq 4$ 
and $\frac{|r_\tau|}{\hat{\sigma}_{\text{bv},\tau}} > 4$ for each $t^{\prime\prime} < \tau < t^\prime$ and for each $t^\prime < \tau < t$.
The value taken for the parameter of the exponentially weighted moving average is $\alpha = \frac{2}{M+1}$, with $M = 60$, corresponding to a half-life time of nearly 21 minutes.


\end{document}